\documentclass[12pt,a4paper]{article}

\usepackage[T1]{fontenc}
\usepackage{multirow,scalerel}
\usepackage{setspace}
\usepackage{verbatimbox}
\usepackage{lscape}
\usepackage{footnote}
\makesavenoteenv{tabular}
\usepackage[numbered]{bookmark}

\usepackage[top=1.5in,
  bottom=1.5in,
  left=1in,
  right=1in]{geometry}

\usepackage{appendix}
\usepackage{amsfonts,enumerate,amsmath}
\usepackage{amsthm}
\usepackage{latexsym}
\usepackage{color}
\usepackage{natbib}
\usepackage{array}
\usepackage{caption}
\usepackage{verbatim}
\usepackage{graphicx}
\usepackage{float,bm}
\usepackage{xcolor}
\usepackage{algorithm}
\usepackage{algpseudocode}
\usepackage{tikz}

\renewcommand{\qedsymbol}{\ensuremath{\Box}}

\graphicspath{{./images/}}

\renewcommand{\cite}{\citet*}

\newcommand{\bqn}{\begin{eqnarray*}}
\newcommand{\eqn}{\end{eqnarray*}}

\newcommand{\bw}{\boldsymbol{\omega}}
\usepackage{chngcntr}
\usepackage{pdfpages}

\newcommand{\bX}{\mathbf{X}}

\newcommand{\bR}{\mathbf{R}}

\newcommand{\bI}{{\bf I}}

\DeclareMathOperator{\diag}{diag}

\newcommand{\bS}{{\bf S}}

\newcommand{\bP}{{\mathbf P}}

\newcommand{\bbeta}{\boldsymbol{\beta}}

\newcommand{\bOmega}{\boldsymbol{\Omega}}
\newcommand{\bomega}{\boldsymbol{\omega}}
\newcommand{\bSigma}{\boldsymbol{\Sigma}}

\numberwithin{equation}{section}

\newtheorem{lem}{Lemma}
\theoremstyle{remark}
\newtheorem{remark}{Remark}
\let\remarkWithoutQED\remark
\let\endremarkWithoutQED\endremark
\renewenvironment{remark}
  {\remarkWithoutQED}
  {\unskip\nobreak\hfill\penalty10000\hbox{\qedsymbol}%
   \endremarkWithoutQED}
\theoremstyle{plain}

\newtheorem{assumption}{Assumption}

\bibpunct[, ]{(}{)}{,}{a}{}{,}%

\global\long\def\one{\boldsymbol{1}}%
\global\long\def\R{\mathbb{R}}%
\global\long\def\DeltaN{\Delta_{N}}%

\providecommand{\propositionname}{Proposition}
\providecommand{\theoremname}{Theorem}
\newtheorem{thm}{\theoremname}
\newtheorem{prop}{\propositionname}

\hypersetup{
  colorlinks = true,
  urlcolor = blue,
  linkcolor = blue,
  citecolor = blue
}

\usepackage{array} \newcolumntype{C}[1]{>{\centering\arraybackslash}p{#1}}

\begin{document}
\title{ Zero Variance Portfolio
}
 \author{Jinyuan Chang\thanks{Southwestern University of Finance and Economics, E-mail: \href{mailto:changjinyuan@swufe.edu.cn}{changjinyuan@swufe.edu.cn}.} \thanks{Academy of Mathematics and Systems Science, Chinese Academy of Sciences.} \and Yi Ding\thanks{ University of Macau, E-mail: \href{mailto:yiding@um.edu.mo}{yiding@um.edu.mo}.}\and Zhentao Shi\thanks{Corresponding Author. The Chinese University of Hong Kong, E-mail: \href{mailto:zhentao.shi@cuhk.edu.hk}{zhentao.shi@cuhk.edu.hk}.  } \and Bo Zhang\thanks{University of Science and Technology of China, E-mail: \href{mailto:wbchpmp@ustc.edu.cn}{wbchpmp@ustc.edu.cn}. } }

\date{}
\maketitle

\begin{abstract}
\noindent

When the number of assets is larger than the sample size, the minimum variance portfolio interpolates the training data, delivering pathological zero in-sample  variance. We show that if the weights of the zero variance portfolio are learned by minimizing the Euclidean norm, it enjoys out-of-sample generalizability on new test data and exhibits the double descent phenomenon. With refinement, it can achieve optimal risk in the over-parametrized regime when the number of assets dominates the sample size. In contrast, if we use the pseudoinverse to deal with the non-invertible sample covariance matrix,  the resulting pseudoinverse weight fails to interpolate the training data and diverges away from the out-of-sample optimality. Extensive simulations and empirical studies demonstrate that the zero variance portfolio performs competitively in high-dimensional portfolio optimization.
\end{abstract}

\vskip 0.3cm

\noindent {Keywords:} Factor Model, Machine Learning, Minimum Variance Portfolio, Random Matrix Theory, Risk Management

\onehalfspacing
\counterwithout{equation}{section}
\section{Introduction}\label{sec:Intro}

Artificial intelligence (AI) and modern machine learning are increasingly used as decision engines in finance, from automated investment management to risk control. A defining feature of these systems is that they are often trained in a data regime that is simultaneously rich in information and scarce in samples. Decision makers can access thousands of correlated signals or assets, yet only a limited time series is available to learn the models. This raises a central issue in AI for finance and business decisions: how to design optimization and estimation procedures that can exploit high-dimensional information while delivering reliable out-of-sample (OOS) performance. 

Recent advancements in AI and machine learning have reshaped how researchers approach return prediction and asset pricing. Emerging financial applications highlight the benefits of complex models and over-parameterization in capturing sophisticated market dynamics (e.g., \citealt{gu2020empirical, fan2022structural, avramov2023machine,chen2024deep}). 
 Empirical evidence shows that highly parameterized models can outperform traditional dimension reduction techniques, even when they interpolate the in-sample data. \cite{kelly2024virtue} show that model complexity benefits the market return prediction, and \cite{Kelly2022} demonstrate that the ``virtue of complexity'' exists for a variety of asset classes. \citet{didisheim2023complexity} provide a theoretical analysis of this phenomenon in overly parameterized asset 
pricing models. Furthermore, in economic predictions, \cite{liao2023does} find that the ridgeless regression that interpolates  in-sample data surpasses many
commonly used models based on dimension reduction.

This counter-intuitive success is closely tied to implicit regularization, where over-parameterized models generalize well despite perfect in-sample fit. Machine learning theory offers insights into this behavior. \cite{Hastie2022} show the
double descent behavior for the ridgeless regression under a non-asymptotic setting. \citet{bartlett2020benign} and \citet{tsigler2023benign} establish the benign of overfitting under the high-dimensional regime where the sample size is of the same order as the number of parameters. \cite{cheng2024dimension} further provide sharp guarantees for benign overfitting under high-dimensional settings where the parameter space far exceeds the sample size.

In this paper, we explore how modern insights from AI and machine learning can benefit portfolio optimization. Since Markowitz (\citeyear{markowitz1952portfolio}) introduced the mean-variance framework, this classic problem has faced persistent challenges that mirror those in modern machine learning. Practitioners operate in a regime that they have limited time series data to construct portfolios that span thousands of assets. 
We focus on the minimum variance portfolio (MVP) problem, which sits at the foundation of modern portfolio theory. 
It serves as both a theoretical cornerstone and a practical default when return estimates are noisy.

\subsection{MVP in High Dimension}

MVP is a simple constrained quadratic optimization. 
It provides the cleanest environment for in-depth analysis. 
Denote $N$ as the number of assets over the cross section, and~$T$ as the sample size over the time dimension. 
Let $\mathbf r_t = (r_{1t},\ldots, r_{Nt})^\top  $ be an $N$-dimensional vector of observed returns (with zero mean) at time $t = 1,\ldots,T$.
We assume that $\mathbf r_t$ is stationary over time, and the population covariance matrix $\boldsymbol{\Sigma} = \mathbb E(\mathbf r_t \mathbf r_t^\top)$ is strictly positive definite. The weight 
$\boldsymbol{\omega} = (\omega_1,\ldots,\omega_N)^\top $ assigned to the $N$ assets must be in the affine hyperplane  
$\DeltaN\;=\;\{\boldsymbol{\omega}\in\R^{N}:\,\one^\top\boldsymbol{\omega}=1\}$,
where $\mathbf 1$ is a column of~$N$ ones. We allow short selling with $\omega_i <0 $. 
Were there an ``oracle'' about $\boldsymbol{\Sigma}$, we would solve the population version of MVP
$$
\min_{\bomega \in \DeltaN} \ \bomega^{\top} \bSigma \bomega.
$$
The oracle weight admits a closed-form solution $\boldsymbol{\omega}^*=\boldsymbol{\Sigma}^{-1}\mathbf{1}/(\mathbf{1}^{\top}\boldsymbol{\Sigma}^{-1}\mathbf{1}),$
which leads to the oracle minimum variance $(\mathbf{1}^{\top}\boldsymbol{\Sigma}^{-1}\mathbf{1})^{-1}>0$. 

In reality $\boldsymbol{\Sigma}$ is unobservable. What the researcher has is an $N\times T$ data matrix $\mathbf{R}=(\mathbf r_{1}, \ldots, \mathbf r_{T})$. A natural and common practice is to replace the population $\boldsymbol{\Sigma}$ with the plain sample covariance matrix 
$\mathbf S=T^{-1}\mathbf{R}\mathbf{R}^{\top}$  to solve
\begin{equation}
    \mathop{\text{min}}_{\boldsymbol{\omega} \in \DeltaN} \ {\boldsymbol{\omega}}^{\top}\mathbf S \, \boldsymbol{\omega}.
\label{eq:mvp}
\end{equation}
When $\mathbf S$ is invertible, which holds in general when $N < T$, the solution is 
\begin{equation}\label{plugin_mvp}
    \widehat{\boldsymbol{\omega}}
    =\frac{\mathbf S^{-1}\mathbf{1}}{\mathbf{1}^{\top}\mathbf S^{-1}\mathbf{1}}.
\end{equation}
In the classical asymptotic framework where $N$ is fixed and $T\to \infty$, under standard assumptions
the law of large numbers ensures that $\mathbf S$ converges in probability to $\boldsymbol{\Sigma}$, thereby with the plug-in weight $\widehat{\boldsymbol \omega}$ converging to the oracle weight $\boldsymbol{\omega}^*$ asymptotically. 

However, the above classical asymptotic framework does not fit the reality of the financial market well. As emphasized above, investors typically need to learn a high-dimensional portfolio based on training data with a relatively small sample size. For instance, if we construct a portfolio that includes about 500 stocks, say the S\&P 500,  and use the widely accessible daily data, then each
month has only about 22 trading days---a short estimation window makes trading strategies adaptive to the evolving market environment. In this case, we are faced with a number of assets $N$ more than twenty  times greater than the sample size $T$. This estimation problem should be studied under a high-dimensional setting where the cross section can be much larger than the sample size.  In the financial world ``$N> T$ '', or more precisely ``$N\gg T$ '', is the norm, rather than the exception.

The MVP literature has long recognized the challenges posed by high dimensionality. 
It is well-known that the plug-in estimator using the sample covariance matrix does not work satisfactorily when the dimension $N$ is high; see, e.g., \citet{michaud1989markowitz, kan2008distribution} and \citet{ao2019approaching}. To address this, various studies have proposed alternative covariance matrix estimators. For example, \cite{caner2025} propose  nodewise regressions for the precision matrix, \cite{caner2024navigating} carry constraints into portfolio analysis,
and \citet{ding2021high} adopt a threshold-type estimator under factor models.
These regularization methods avoid the plain sample covariance matrix $\mathbf S$ in high dimensions.

Ridge-type regularization is one of the off-the-shelf choices, and there is growing interest in it in portfolio optimization problems  (e.g., \citealt{bodnar2018estimation, bodnar2022optimal, bodnar2024two}, and \citealt{meng2025estimation}). Consider \begin{equation}
    \mathop{\text{min}}_{\boldsymbol{\omega} \in \DeltaN} \  \{{\boldsymbol{\omega}}^{\top}\mathbf S \, \boldsymbol{\omega} + \tau \| \boldsymbol{\omega} \|^2 \},
\label{eq:mvp-ridgelet}
\end{equation}
which is solved by 
\begin{align}\label{eq:ridgelet_w}
\widehat{\boldsymbol{\omega}}_{\tau}=
\frac{\mathbf{S}_\tau^{-1}\mathbf{1}}{\mathbf{1}^{\top}\mathbf{S}_\tau^{-1}\mathbf{1}}~~\textrm{with}~~\mathbf{S}_\tau=\mathbf S+\tau \mathbf{I}_N.
\end{align}
Here $\mathbf{I}_N$ is the identity matrix.
Ridge uses $\tau>0$ as a tuning parameter to balance the bias and variance. In practice, the tuning parameter is chosen either by cross validation from a grid of user-specified values, or using \cite{ledoit2004well}'s \emph{bona fide} linear shrinkage. 
Notice that \eqref{plugin_mvp} is the limit of \eqref{eq:ridgelet_w} when $\tau \to 0$ if $\mathbf S$ is invertible.

In the standard framework of the bias-variance tradeoff, overfitting fails to generalize to a test dataset, leading to poor OOS performance. 
In recent years, however, this piece of conventional wisdom has been challenged when AI researchers witnessed that 
deep neural networks with many more parameters than training examples were achieving state-of-the-art generalization and that larger models continued to deliver better test performance.
To describe the interactions between generalizability and model complexity, \citet{belkin2020two} use the term \emph{double descent} to refer to the phenomenon where the test error of a model displays two distinct regimes of descent as model complexity increases. 
In regressions, double descent has been comprehensively studied by \cite{Hastie2022} via the ridgeless estimator, which is the interpolation solution.

Inspired by \cite{Hastie2022} and \cite{shi2025}, this paper studies the minimum-$\ell_2$-norm weight in MVP when $N>T$:
\begin{equation}\label{eq:P0}
\min_{\boldsymbol{\omega}\in \DeltaN}\left\Vert \boldsymbol{\omega}\right\Vert ^{2}\qquad\text{subject to } \boldsymbol{\omega}^{\top} \mathbf S \boldsymbol{\omega} = 0.
\end{equation}
We call it the \emph{zero variance portfolio} (ZVP) problem because the in-sample variance $\boldsymbol{\omega}^{\top} \mathbf S \boldsymbol{\omega} = 0$.
Conventionally, zero in-sample variance is viewed as a pathological case because it drops below the oracle population variance $(\mathbf{1}^{\top}\boldsymbol{\Sigma}^{-1}\mathbf{1})^{-1}>0$, and it is equivalent to $\bR^\top \bomega = \mathbf 0$, which interpolates, or \emph{overfits}, the training data. 
Lemma \ref{lem:existence-of-zero-var} in Section \ref{sec:exact_zvp} shows that the solution to \eqref{eq:P0} exists in general when $N>T$, and admits a closed-form
\begin{equation}\label{eq:ZVP_w}
\widehat{\boldsymbol{\omega}} = \frac{\bP^\perp_S \one }{\one^\top \bP^\perp_S \one},      
\end{equation}
where $\bP^\perp_S$ is the projector to the null space of $\mathbf S$ (formally defined in Section \ref{sec:exact_zvp}).  
When $N>T$, we will see in Proposition \ref{prop:ridgelet} in Section \ref{sec:exact_zvp}  that the ZVP estimator \eqref{eq:ZVP_w} is the limiting case of the ridge weight $\widehat \bomega_\tau$ as $\tau \to 0$. On the other hand, when $N\leq T$, the ridge weight $\widehat{\bomega}_\tau$ converges to the plug-in weight $\widehat{\bomega}$ given in \eqref{plugin_mvp} as $\tau\rightarrow0$, provided that $\mathbf S$ is invertible. To unify the notation for $N\leq T$ and $N>T$, we reformulate $\widehat{\bomega}$ as follows:
\begin{equation}
    \label{eq:zvp_both}
\widehat{\bomega} :=
\begin{cases}
\displaystyle\frac{\mathbf S^{-1}\mathbf{1}}{\mathbf{1}^{\top}\mathbf S^{-1}\mathbf{1}}, &\text{when} \   N\leq T \ \text{and} \ \bS\  \text{is invertible},\\[7pt]
\displaystyle \frac{\mathbf P^\perp_S \one }{\one^\top  \mathbf P^\perp_S \one}, &\text{when} \    N > T.
\end{cases}
\end{equation}
With a slight abuse of terminology, in the sequel
we will refer to $\widehat{\bomega}$ given in \eqref{eq:zvp_both} as the \emph{ZVP weight}, or ZVP for simplicity.

\begin{figure}[htbp]
\begin{center}
\includegraphics[page=1,width=0.49\textwidth]
{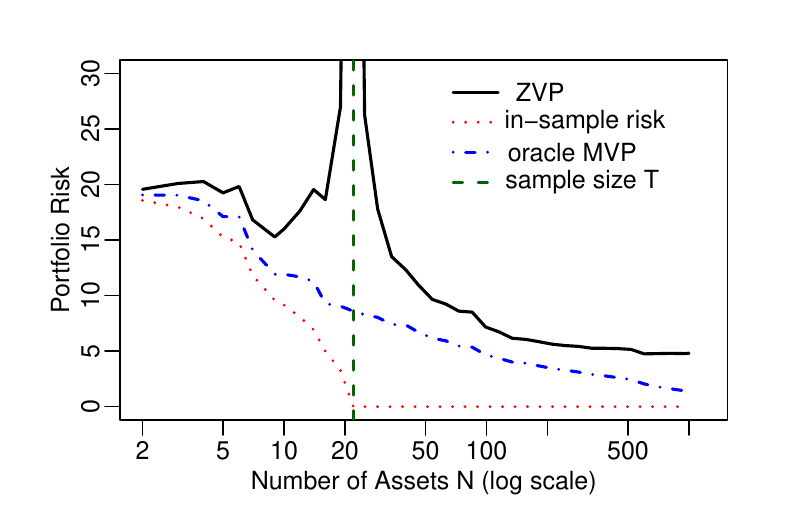}
\includegraphics[page=2,width=0.49\textwidth]
{Figs/simu_setting2_augdoubledescent_absrisk_illustrate_ExactZVP}

\caption{{\textbf{Risk curve of the MVP estimated by $\widehat{\bomega}$ in \eqref{eq:zvp_both}  (left) and the pseudoinverse estimator (right). } The oracle minimum risk is a benchmark. Risk is measured by the standard deviation of portfolio returns. 
The returns are generated from a factor model described in Section \ref{sec:simu-DGP}. Specifically, we use Setting 2 of the simulation design, in which the idiosyncratic covariance matrix is sparse.  }
}
\label{dd_risk}
\end{center}
\end{figure}

The portfolio based on the ZVP weight $\widehat{\bomega}$ in \eqref{eq:zvp_both} features the double descent in its risks, measured by the standard deviation of portfolio returns, 
which is indicated by its regime shift in terms of the relative size of $N$ and $T$. 
We illustrate such a pattern in the left panel of Figure \ref{dd_risk}. 
We draw the portfolio risk (black solid curve) against  the number of assets~$N$  while keeping the sample size fixed at $T=22$.
The portfolio OOS risk first decays
as $N$ grows from  2 to~10, that is, the first descent. Then it rapidly grows as $N$ approaches~$T$. As $N$ goes beyond~$T$, the risk dips again in a second descent.  
On the other hand, when $N>T$,  the in-sample risk (red dotted curve) lies flat at  zero.

Some readers might be surprised that the pseudoinverse $\mathbf S^+$ does not show up in the estimated weight in the over-parameterized regime $N>T$. 
In linear regressions, it is well-known that the limiting ridge estimator replaces the inverse~$(\cdot)^{-1}$ of the Gram matrix in the usual OLS estimator with the pseudoinverse $(\cdot)^{+}$.
For comparison, let us mimic the ridgeless regression estimator by considering 
\begin{equation}\label{eq:ridgeless_w}
\widehat{\bomega}_+=
\frac{\mathbf S^{+}\mathbf{1}}{\mathbf{1}^{\top}\mathbf S^{+}\mathbf{1}}
\end{equation}
when $\mathbf{1}^{\top}\mathbf S^{+}\mathbf{1}>0$.
We call $\widehat{\bomega}_+$ the \emph{pseudoinverse weight} in this paper.\footnote{A note about the terminologies. Technically, $\widehat \bomega$ in \eqref{eq:zvp_both} is the \emph{ridgeless MVP}.
However, the ridgeless MVP and the \emph{ridgeless regression} are two distinct problems, where the latter is associated with the pseudoinverse. To avoid confusion, when $N>T$ we call $\widehat \bomega$ the ZVP weight, and $\widehat{\bomega}_+$ the pseudoinverse weight.
}

In the MVP problem, $\widehat{\bomega}$ and $\widehat{\boldsymbol{\omega}}_+$ are fundamentally different estimators.
To see intuitively why the pseudoinverse is undesirable,
in the right panel of Figure~\ref{dd_risk} we draw its risk with the same data as in the left panel. 
When~$N<T$, the OOS risk profile of the pseudoinverse weight (black solid curve) is the same as that of $\widehat{\bomega}$.
However, after a quick second descent when $N$ exceeds $T$, the risk with the pseudoinverse diverges from the oracle counterpart when $N$ becomes larger than~50. 
Moreover, when $N>T$, the in-sample risk (red dotted curve) rises as well---it does not interpolate the training data.

\subsection{Refining ZVP}

In real-world financial markets, stock returns are driven by common factors at the overall market level, as well as correlations at a localized level from industries and supply chains. A low-rank factor structure with a sparse idiosyncratic covariance matrix is commonly used to characterize the comovements (e.g., \citealt{ding2021high,du2023high}). 
To better absorb information from the local level, we propose a refined version of ZVP. 
The refined ZVP estimator further leverages an estimator of the covariance matrix of the idiosyncratic component. To differentiate the two alternative ZVP estimators, we refer to $\widehat \bomega$ in \eqref{eq:zvp_both} as \emph{ZVP1} and the refined version as \emph{ZVP2} (denoted as $\widetilde{\bomega}$ in \eqref{eq:ridgelet2_w} in Section \ref{sec:ridgelet2}).\footnote{These names are inspired by ``System~1'' and ``System~2'' in the popular science book \textit{Thinking, Fast and Slow} by Daniel Kahneman, a psychologist and Nobel laureate in economics.}
We show that when  $N\gg T$, ZVP2 improves upon ZVP1 by attaining optimality if the stock returns follow an approximate factor model \citep{chamberlain1983arbitrage}.
Here ``optimality'' is defined in the sense that the OOS risk of the portfolio constructed by the weights from ZVP2 asymptotically approaches the risk of the population oracle (see Theorem \ref{t1s} in Section \ref{sec:ridgelet2}).
It benefits from the \emph{blessing of  dimensionality} \citep{li2018embracing}.

\begin{figure}
\centering{}\includegraphics[page=3,width=0.7\textwidth]{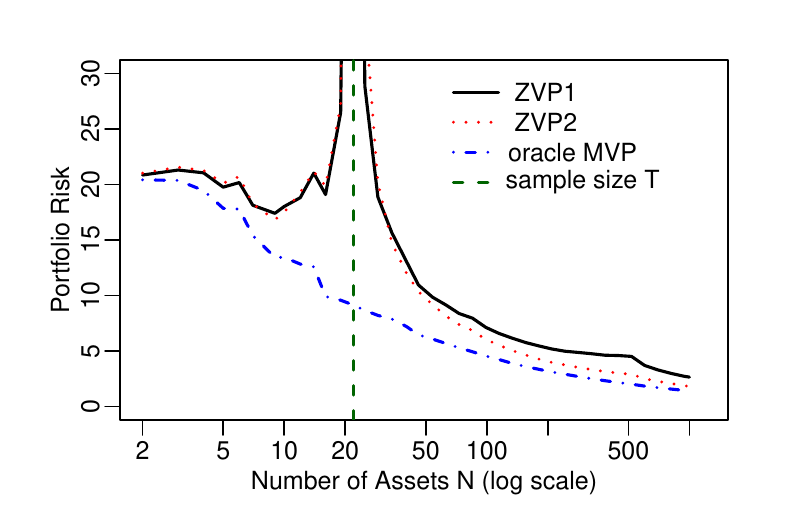}
\caption{{\bf {Risk curve of the MVP estimated with ZVP1 and ZVP2}.} Risk is measured by the standard deviation of portfolio returns. We use a setting with factor model and sparse idiosyncratic covariance matrix. Specifically, we use the same factor loadings as those specified in Section \ref{sec:simu-DGP} and a sparse idiosyncratic covariance matrix with correlation: $\rho_{i,j} = 0.8^{|i-j|}$ when $|i-j|\leq 10$, and 0 otherwise. 
}\label{ridgelet2}
\end{figure}

To witness the benefit of ZVP2 and its improvement over ZVP1, in Figure \ref{ridgelet2} we show the risks of the two ZVP estimators. 
Here, we consider a factor model setting with a sparse idiosyncratic covariance matrix.   ZVP2 yields a visually lower risk for large $N$, and its performance approaches the oracle minimum risk as $N$ increases.

\subsection{Contributions and Outline}
Our theoretical contributions are twofold. First, we show that ZVP1 generalizes to new test data after passing the peak of the complexity regime, whereas the pseudoinverse estimator fails to achieve in-sample interpolation.
Second, in the approximate factor model, we make it clear that optimality hinges on knowledge of the idiosyncratic covariance matrix, and by leveraging a consistent estimator of it, ZVP2 can deliver optimality in the high-dimensional case when $N\gg T$. ZVP2 is our recommended estimator.

Our asymptotic analysis relies on Random Matrix Theory (RMT). 
RMT is a principled toolkit for studying eigenvalue behavior and other high-dimensional phenomena in AI and machine learning, and has seen 
increasing adoption; see, e.g., \citet{onatski2018alternative, Kelly2022, bykhovskaya2023high}, and \citet{he2024ridge}.

We perform extensive numerical studies to evaluate the performance of the proposed ZVP approach. We compare it with several benchmark methods, including the linear shrinkage (LS) estimator (\citealp{ledoit2004well}) and the factor-based nonlinear shrinkage (FNLS) estimator (\citealp{ledoit2017nonlinear}). Using the S\&P 500 Index constituents and Nikkei 225 stocks, the ZVP estimators perform competitively.
ZVP allows us to assign weights to more than 700 stocks with merely one month's daily returns. 

Our paper adds to the recent debate about the ``virtue of complexity'' \citep{berk2023comment, nagel2025seemingly, kelly2025understanding}. While these papers mostly focus on the complexity of representation in a regression context, our paper provides theoretical justifications and empirical evidence that expanding the space of accessible assets, together with a proper, simple estimation strategy, can reduce portfolio risks despite relatively small sample sizes. It also complements the latest works in empirical finance using sophisticated, tailored, and nonlinear machine learning algorithms such as the sparse Bayesian clustering \citep{CongFengHeLi2023SparseModeling},
panel trees \citep{CongFengHeHe2025PanelTrees}, deep reinforcement learning and generative models \citep{CongTangWang2026AlphaPortfolio, CampelloCongZhou2026AlphaManager} in predictive frameworks with large datasets with rich features. The MVP problem in our setting is linear and amenable to theoretical analysis.

The rest of the paper is organized as follows. Section \ref{Sec:thm} constructs the ZVP estimators and presents the main theoretical results. Sections \ref{Sec:Simu} and \ref{Sec:Emp} provide simulation studies and empirical applications, respectively. 
Appendix \ref{Proofs} includes the proofs of Lemma \ref{lem:existence-of-zero-var} and Proposition \ref{prop:ridgelet}. 
Long proofs of the remaining theoretical statements and additional numerical results are collected in the online supplementary material.

\section{Theory}\label{Sec:thm}

We use the following notations throughout the paper.  A vector is a column by default. For a generic vector $\mathbf{x}=(x_i)$, we define its transpose as $\mathbf x^\top$, and its $\ell_2$-norm as $\|\mathbf{x}\|=\sqrt{\sum_{i}x_i^2}$.
For a generic matrix $\mathbf{A}$,
we denote its spectral norm as $\| \mathbf{A}\|=\max_{\|\mathbf{x}\|\leq 1}\|\mathbf{A}\mathbf{x}\|$, its minimum singular value  as $\|\mathbf{A}\|_{\min}$, its rank as ${\rm rank}(\mathbf{A})$, 
and its trace as $\mathrm{tr}(\mathbf A)$ when $\mathbf{A}$ is a square matrix.
We use ``$\stackrel{{\rm p}}{\to }$'' to represent
convergence in probability. For two sequences of positive real numbers $\{a_n\}$ and $\{b_n\}$, we write $a_n \asymp b_n$ if there exists some constant $c\geq 1$ such that $b_n/c \leq a_n \leq c b_n$ for any $n\geq 1$, and write $a_n=o(b_n)$ or $a_n\ll b_n$ if $\limsup_{n\rightarrow\infty}a_n/b_n= 0$.

\subsection{ZVP and Ridge}\label{sec:exact_zvp}

We first present some linear algebra properties about the estimators under consideration when $N>T$. Recall $\mathbf R$ is an $N\times T$ data matrix. If ${\rm rank}(\mathbf R)= T$, we diagonalize $\mathbf S$ as
\begin{equation}\label{eq:svd_s}
\mathbf S = {\mathbf U}_T {\mathbf \Lambda}_T {\mathbf U}^\top _T,    
\end{equation}
where the diagonal matrix ${\mathbf\Lambda}_T$ stores the $T$ nonzero eigenvalues of $\mathbf S$, and the columns of the $N\times T$ matrix $\mathbf U_T$ are the corresponding eigenvectors. The condition ${\rm rank}(\bR)=T$ can be easily satisfied in practice when $N>T$. 
The projector onto the null space of $\mathbf S$ is $\mathbf P^\perp_S=\mathbf I_N-\mathbf U_T\mathbf U_T^\top$. 

\begin{lem}\label{lem:existence-of-zero-var}
Suppose $N >T $ and  $\one \notin \mathrm{span}(\mathbf R)$. Then $\widehat{\bomega} = \bP^\perp_S \one / (\one^\top \bP^\perp_S \one)$ is the unique solution to the ZVP problem \eqref{eq:P0}.
\end{lem}

Throughout this section, we work with $N>T$ and assume $\one \notin \mathrm{span}(\mathbf R)$. The one-vector cannot be a linear combination of the columns of $\mathbf R$;
otherwise the denominator of $\widehat{\bomega}$ becomes $\one^\top \bP^\perp_S \one = 0$ and is thus not well-defined. This condition generally holds with probability one when the entries of $\mathbf R$ are random, under which  
Lemma \ref{lem:existence-of-zero-var} confirms the existence of ZVP and its closed-form.

Next, we turn to the ridge weight $\widehat{\bomega}_\tau$ in \eqref{eq:ridgelet_w}. 
The following proposition clarifies the relationship between the ZVP estimator and the ridge counterpart.

\begin{prop}\label{prop:ridgelet}
Under the conditions in Lemma \ref{lem:existence-of-zero-var} and ${\rm rank}(\mathbf R)= T$, the inverse of the matrix $\bS_\tau$ defined in \eqref{eq:ridgelet_w} can be formulated as 
\begin{equation}
\bS_\tau^{-1}
=\tau^{-1}\mathbf P_S^\perp
+\mathbf U_T(\mathbf\Lambda_T+\tau\mathbf I_T)^{-1}\mathbf U_T^\top 
\label{eq:smw-ridge}
\end{equation}
for any $\tau>0$, 
and thus the ridge weight can be written as
\begin{equation}\label{Ridgelet_repres}
\widehat{\boldsymbol{\omega}}_{\tau}
= \frac{\mathbf P_S^\perp\one+
\tau\mathbf U_T(\mathbf\Lambda_T+\tau\mathbf I_T)^{-1}\mathbf U_T^\top\one}
{\one^\top\mathbf P_S^\perp\one+
\tau\one^\top\mathbf U_T(\mathbf\Lambda_T+\tau\mathbf I_T)^{-1}\mathbf U_T^\top\one}.
\end{equation}
It follows that 
$
\widehat{\bomega}_\tau \rightarrow \widehat{\boldsymbol{\omega}}
$ 
as $\tau \to 0$.
\end{prop}

Equation \eqref{eq:smw-ridge} shows that the inverse of $\bS_\tau$ has two components: the first term $\tau^{-1}\mathbf P_S^\perp$ in the null space of $\mathbf S$ and the second term 
$\mathbf U_T(\mathbf\Lambda_T+\tau\mathbf I_T)^{-1}\mathbf U_T^\top$
in the column space of $\mathbf S$.
When $\tau \to 0$, the first term will dominate the second term, leading to a divergent $\bS_\tau^{-1}$. 
Nevertheless, $\widehat{\bomega}_\tau$ has a stable limit as $\tau \to 0$, because the denominator in \eqref{Ridgelet_repres} standardizes the scale and cancels out the divergence in the numerator. In the limit, the second, dominated  component vanishes, and only the terms involving $\bP_S^\perp$ remain. 
The last statement in Proposition \ref{prop:ridgelet} confirms that the ZVP estimator is the limiting ridge weight.

\begin{remark}\label{remark:bodnar}
The MVP with ridge penalty has been extensively studied by \cite{bodnar2018estimation, bodnar2022optimal, bodnar2024two}. When $N>T$,  \cite{bodnar2022optimal}  discuss the small ridge tuning parameter~\mbox{$\tau \to 0$} around the peak when $N/T \to 1$ as a device to smooth out the discontinuity between~$\mathbf S^{-1}$ and $\mathbf S^+$. 
Our paper distinguishes itself from them in the following aspects. First, these papers take the ridge penalty level $\tau$ as a tuning parameter to be optimized; we instead work with the ZVP estimator, which is tuning-free. Second, while these papers consider~$N/T \to \gamma \in (0,\infty) $, we encourage a very large $N$ with $N\gg T$  to obtain the optimality of the ZVP approach (see Section \ref{sec:ridgelet2}). Moreover, we show that, when $N\gg T$, our ZVP achieves the same asymptotic out-of-sample risk as the optimally tuned ridge portfolio (see Proposition \ref{tauoptimal}).
\end{remark}

\subsection{Why Not Pseudoinverse}\label{sec:ridgelet}

Substituting \eqref{eq:svd_s} into the closed-form of the pseudoinverse weight \eqref{eq:ridgeless_w} yields
\begin{equation}\label{eq:Ridgeless_w}
\widehat{\boldsymbol{\omega}}_+ = 
\frac{\mathbf S^{+}\mathbf{1}}{\mathbf{1}^{\top}\mathbf S^{+}\mathbf{1}}
= \frac{ \mathbf U_T \mathbf \Lambda_T^{-1} \mathbf U_T^\top \one}{\one^\top \mathbf U_T  \mathbf \Lambda_T^{-1} \mathbf U_T^{\top} \one}.
\end{equation}
Clearly, it lies in the column space of $\mathbf S$, whereas the ZVP weight in \eqref{eq:ZVP_w} lies in the null space of $\mathbf S$. These two estimators are actually orthogonal: $\widehat{\boldsymbol{\omega}}^{\top}\widehat{\boldsymbol{\omega}}_+=0$ as \mbox{$\mathbf P^\perp_S\mathbf U_T=\mathbf 0$}. Moreover, in-sample interpolation cannot be achieved by $\widehat \bomega_+$, because 
$
\widehat \bomega_+^\top \mathbf S \widehat \bomega_+  = 1/(\one^\top \mathbf S^+ \one)> 0.
$

\begin{remark}
The role of the pseudoinverse is fundamentally different in  linear regression and in MVP. 
MVP is an unsupervised learning problem, with the data matrix $\bR$ only. Linear regression, instead, is a supervised learning problem, with an additional response vector $\mathbf y\in\mathbb{R}^T$. 
To be more directly comparable with our setting, we denote  $\bR$ as the design matrix.
When $N>T$, the interpolation system $\bR^\top \boldsymbol{\beta}=\mathbf y$ is feasible. The ridge regression estimator is\footnote{If we use the usual notation $\bX$ ($T\times N$ matrix) as the design matrix, by setting $\bX = \bR^\top$ we recover the common expression of the ridge regression estimator
$\widehat \bbeta_\tau = (\bX^\top \bX + \tau_T \bI_N)^{-1} \bX^\top \mathbf y$, where $\tau_T = T\tau$.
}
\begin{align*}
\widehat{\boldsymbol{\beta}}_\tau & 
=(T^{-1}\bR \bR^\top+ \tau\mathbf I_N )^{-1} \bR \mathbf{y}/T = 
\bS_\tau^{-1} \bR \mathbf{y}/T \notag \\
& = [\tau^{-1}\mathbf P_S^\perp
+\mathbf U_T(\mathbf\Lambda_T+\tau\mathbf I_T)^{-1}\mathbf U_T^\top] \bR \mathbf y / T \notag \\
& = \mathbf U_T(\mathbf\Lambda_T+\tau\mathbf I_T)^{-1}\mathbf U_T^\top \bR \mathbf y / T \\ 
&\to \mathbf S^+ \bR \mathbf y / T 
\end{align*}
as $\tau\rightarrow0$, where the second line follows by \eqref{eq:smw-ridge} and 
the third line by the fact $\bP_S^\perp \bR = \mathbf 0 $ because $\bR\in\mathrm{span}({\mathbf U}_T)$. 
The post-multiplication by $\bR$ in the second line kills the null space component $\tau^{-1} \bP_S^\perp$ while maintaining $\mathbf U_T(\mathbf\Lambda_T+\tau\mathbf I_T)^{-1}\mathbf U_T^\top$.
This is the key reason why the pseudoinverse appears in the ridge regression estimator; here the null space component is irrelevant. 
\end{remark}

Following the discussion below Proposition \ref{prop:ridgelet}, the ZVP weight $\widehat{\bomega}$ has a different right-hand side; with $\bS_\tau^{-1}$ post-multiplied by $\one$, the null space component $\tau^{-1} \bP_S^\perp$ cannot be eliminated and it plays a dominant role when $\tau \to 0$. In contrast, the pseudoinverse weight $\widehat{\bomega}_+$ lies in the column space of $\mathbf S$ and thus is orthogonal to $\widehat{\bomega}$.
The phenomenon here echoes \citet[p.658]{ChoiKwonLiao2026} when they study the low-rank projections. Their \emph{Tikhonov-regularization function} is discontinuous in a neighborhood of zero and does not converge to the pseudoinverse counterpart either, which is analogous to our $\|\bS_\tau^{-1}\| \to \infty$ as $\tau \to 0$. With proper normalization, their \emph{rescaled Tikhonov-regularization function} has a finite limit, which is again analogous to our stable limit of $\widehat{\bomega}_\tau$.

\begin{remark}
This paper studies MVP. One natural question is whether the limiting approach by setting up a ridge-regularized covariance matrix and then taking $\tau\to 0$
can be extended to mean-variance (M-V) portfolio optimization, which aims at maximizing the Sharpe ratio. Here we highlight a few key distinctions between MVP and M-V portfolio optimization. First, an M-V portfolio needs to satisfy a given risk constraint. It is difficult to achieve the risk target using purely in-sample data  when $N\gg T$. 
Second and more critically, unlike MVP, the contribution of the Sharpe ratio for an M-V portfolio comes from both the factor component and idiosyncratic component (if there is some alpha). In particular, the factor component plays an important role in the optimal Sharpe ratio. For example, \cite{lu2024double} study M-V optimization  under the case when the mean of the stock return is spanned by the factors only. In this case, they argue that one should use the pseudoinverse. Due to these fundamental differences, the ZVP analysis does not directly carry over to M-V optimization. A systematic theoretical analysis would require extending the meticulous calculation of RMT into the M-V setting, which deserves a standalone paper in future research.
\end{remark}

\medskip

\subsection{Out-of-Sample Variance}\label{subsec:OOSV_RISK}

Given that ZVP interpolates the in-sample data when $N>T$, how does it behave in a test dataset? We will investigate this question 
in an asymptotic framework where both $N$ and $T$ grow to infinity.
Define  $\mathbb{V}: \DeltaN \mapsto \mathbb R^+$ as the \emph{OOS variance} of a generic weight $\boldsymbol{\omega}$, given by
$$\mathbb V(\boldsymbol{\omega}) = \boldsymbol{\omega}^\top \boldsymbol{\Sigma} \boldsymbol{\omega}.$$ The oracle OOS variance is 
$\mathbb V(\boldsymbol{\omega}^* ) = (\one^\top \boldsymbol{\Sigma}^{-1} \one)^{-1}$,
which gets smaller when  $N$ is larger.
To measure how far $\mathbb V(\boldsymbol{\omega})$ deviates from its oracle counterpart, 
we define the (OOS) \emph{relative variance} as
$$\mathbb{RV}(\boldsymbol{\omega}) = \frac{\mathbb V(\boldsymbol{\omega} ) }{ \mathbb V(\boldsymbol{\omega}^* )} = \boldsymbol{\omega}^\top \boldsymbol{\Sigma} \boldsymbol{\omega} \cdot \one^\top \boldsymbol{\Sigma}^{-1}\one.$$
Since no estimator can have lower OOS variance than the oracle,  $\mathbb{RV}(\boldsymbol{\omega})\geq 1 $.

\begin{figure}[htbp]
\begin{center}
\includegraphics[width=0.6\linewidth]{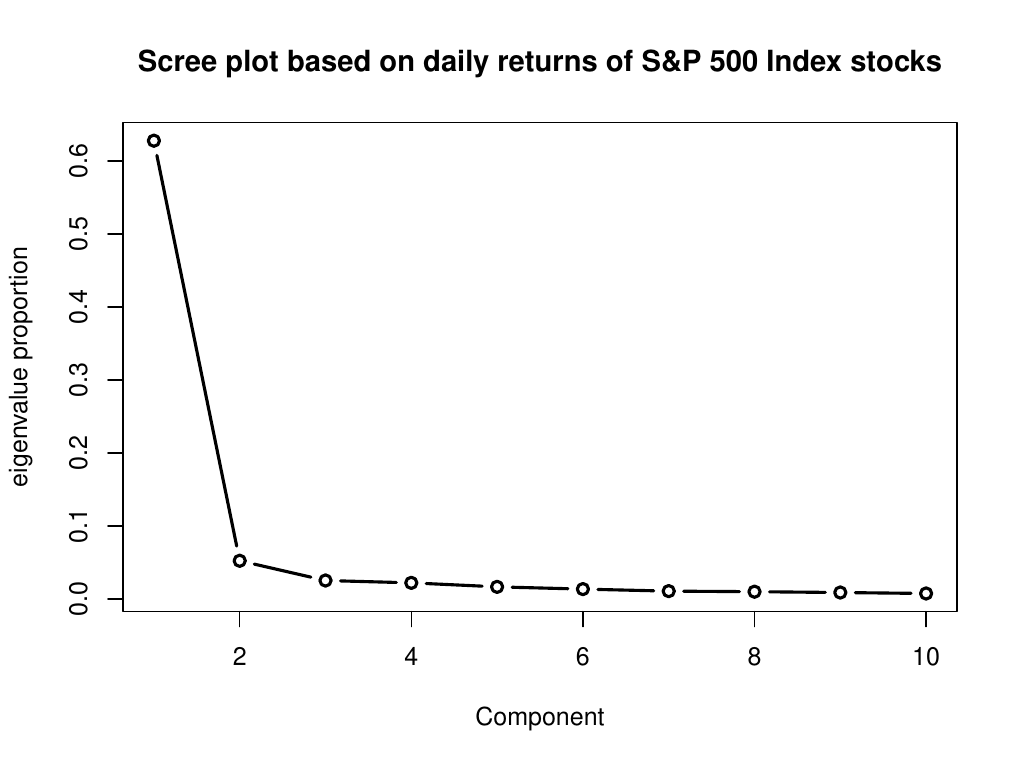}
\caption{\textbf{Scree plot of daily returns of S\&P 500 Index Constituent stocks.} We compute the sample covariance matrix of daily returns of S\&P 500 Index stocks between 2020 and 2023. The Y-axis shows the ratios of its principal eigenvalues over the sum of all eigenvalues.   }
\label{pca_stocks}
\end{center}
\end{figure}

Factor models are fundamentally important for financial returns, as supported by numerous empirical studies. 
Figure \ref{pca_stocks} plots the top 10 eigenvalues of the sample covariance matrix of the S\&P 500 constituent stocks. The first sample eigenvalue clearly dominates all others in  magnitude. 
Suppose that the stock return data matrix $\mathbf R$ follows a latent factor model of $r$ factors
$$\mathbf{R}=\mathbf{BF}+\mathbf{E},$$ where an $N\times T $ matrix $\mathbf E$ represents the idiosyncratic returns, and it is independent of an $r \times T$ matrix of factors $\mathbf{F} = (\mathbf f_1,\ldots,\mathbf f_T) $.
The factor $\mathbf f_t$ is stationary over time, and as a normalization,~\mbox{$\mathbb E( \mathbf  f_t \mathbf f_t^\top )= \mathbf I_r$} and $\|T^{-1}\mathbf{F}\mathbf{F}^\top-\mathbf I_r\|=O_{\rm p}(T^{-1/2})$ which holds under some common mixing conditions. The associated factor loadings are encoded in an $N \times r$  non-random matrix~$\mathbf{B}$. Decompose 
$$
\mathbf B \mathbf B^\top = \mathbf V \mathbf{\Lambda}_{B}\mathbf V ^\top=\mathbf V \ \mathrm{diag}(\lambda_{B,1},\ldots,\lambda_{B,r})  \mathbf V ^\top,
$$
where  $\lambda_{B,1} \geq \cdots \geq \lambda_{B,r} > 0$ are the $r$ eigenvalues and the $N\times r$ matrix  $\mathbf V$ consists of the corresponding eigenvectors. Throughout the theoretical analysis, the number of factors $r$ is fixed and does not depend on $N$
or $T$.
We impose the following assumptions to facilitate asymptotic analysis. 

\begin{assumption}\label{assu}\mbox{ }\\
\vspace{-0.5cm}
\begin{enumerate}
\renewcommand{\labelenumi}{(\roman{enumi}).}
    \item Let $\mathbf{Z}$ be an $N\times T$ matrix consisting of i.i.d.~standard Gaussian entries, and the idiosyncratic returns $\mathbf{E} = \mathbf{\Omega}^{1/2}\mathbf{Z}$,  where 
    $\| \mathbf \Omega\| \asymp \| \mathbf \Omega\|_{\min}  \asymp 1$.
    \item  $ \lambda_{B,1} \asymp \lambda_{B,r}\asymp N^{\delta}$ for some $\delta \in (0,1]$, and $N^{\delta/2} \gg 1+ N / T$.
\end{enumerate}
\end{assumption}

Assumption \ref{assu}(i) requires that the idiosyncratic returns $\mathbf E$ are linear transformations of i.i.d.~entries of $\mathbf Z$. 
Admittedly, Gaussianity is a strong assumption. It is a technical condition to keep the theory simple when RMT is invoked. It is possible to relax Gaussianity by imposing some finite moment conditions instead, at the cost of substantially complicating the notation and proofs but adding little insight into the theoretical mechanisms. There exist methods--—albeit with cumbersome procedures—--in RMT to extend the Gaussianity assumption to general distributions.
For example, \cite{KY2017} relax the conditions so that any finite moment is bounded. \cite{DY2018}  relax the conditions to the fourth moment tail condition and diagonal $\mathbf\Omega$, while \cite{Yang2019} removes the diagonal restriction under the same tail condition.\footnote{In our Monte Carlo simulations, we check the performance of the portfolio risks under both  normal errors and heavy-tailed errors, and the results are robust.} Assumption~\ref{assu}(ii) assumes that the $r$ eigenvalues have the same order of strength for simplicity, and the divergence of the factor strength separates the signal from the idiosyncratic returns in the approximate factor model. The divergence speed of the factor strength, which surely holds if $N \asymp T$, allows relatively weak factors. 

Under Assumption \ref{assu}, the population covariance matrix of $\mathbf r_t$ is
$$
\mathbf{\Sigma} = \mathbf{B} \mathbf{B}^\top+\mathbf{\Omega},
$$
with $r$ spiked eigenvalues of order $N^\delta$, and $(N-r)$ non-spiked eigenvalues of order~1.
By the Sherman-Morrison-Woodbury formula \citep[Equation (0.7.4.1)]{horn2012matrix}, the oracle OOS variance is  
\begin{align*}
\mathbb{V}({\boldsymbol{\omega}}^*)=\frac{1}{\mathbf 1^\top \mathbf \Omega^{-1/2}(\mathbf I_N -\mathbf Q_{\mathbf \Omega^{-1/2} \mathbf V \mathbf{\Lambda}^{1/2}_B} )\mathbf \Omega^{-1/2}\mathbf 1}, 
\end{align*}
with 
$$\mathbf Q_{\mathbf \Omega^{-1/2} \mathbf V \mathbf{\Lambda}^{1/2}_B}=\mathbf \Omega^{-1/2} \mathbf V \mathbf{\Lambda}^{1/2}_B (\mathbf I_r + \mathbf{\Lambda}^{1/2}_B\mathbf V^\top \mathbf{\Omega}^{-1} \mathbf V \mathbf{\Lambda}^{1/2}_B)^{-1}\mathbf{\Lambda}^{1/2}_B\mathbf V^\top \mathbf \Omega^{-1/2}.$$
Clearly, it involves $\mathbf V$ and $\mathbf{\Lambda}_B$ from the factor component, as well as $\mathbf \Omega$ from the idiosyncratic component.
Define $\mathbf P_{\mathbf V}^{\perp}=\mathbf I_N- \mathbf V \mathbf V^{\top}$ as the projector to the null space of $\mathbf{B}^\top$.
The next assumption rules out the unlikely cases that the one-vector
$\mathbf 1$
is in the $r$-dimensional linear space spanned by $\mathbf V$.

\begin{assumption}\label{assu:order} 
$\|\mathbf P_{\mathbf V}^{\perp} \mathbf 1\|/\sqrt N\gg T^{-1}N^{1-\delta/2}$.
\end{assumption}

Since $\sqrt N$ is the $\ell_2$-norm of $\one$, geometrically  $N^{-1/2}\|\mathbf P_{\mathbf V}^{\perp} \mathbf 1\|$ is the $\sin(\cdot )$ of the angle between the one-vector and the linear space spanned by the columns of $\mathbf V$. Since Assumption \ref{assu}(ii) pushes $N^{1-\delta/2}/T \to 0$,  Assumption \ref{assu:order} trivially holds if the angle is non-degenerate.
This assumption is by no means restrictive. {It excludes cases in which the one-vector lies in the loading space or the angle between the two tends to zero too quickly. The angle may be arbitrarily small and may even tend to zero, provided that its sine dominates $T^{-1}N^{1-\delta/2}$.} 

Under Assumptions \ref{assu}-\ref{assu:order} and $T=O(N)$, the MVP can
diversify the overall risk (not only the idiosyncratic one) to zero asymptotically. We provide an illustrative example in Section S.18 of the supplementary material to demonstrate this risk diversification property.
The risk diversification condition has been discussed in \cite{ding2021high}. We impose Assumption~\ref{assu:order} not because the estimator fails to work when it is violated. Instead, it makes the analysis interesting and meaningful.

\begin{remark}
A commonly used industrial practice is the MSCI Barra risk model,  which assumes that
 risk factor loadings are common to all assets. Under this model, \mbox{$\one \in \mathrm{span}(\mathbf V)$}. In this case,  the \emph{systematic risk} from the factors cannot be diversified, and the order of~$\mathbb V(\boldsymbol{\omega}^*)$ is $N^{\delta-1}$. Hence, the risk of equal weight portfolio would be of the same order as the oracle minimum risk, while the effect of $\mathbf \Omega$ becomes negligible in the first-order asymptotics. As a result, this setting can be covered by \citet{shi2025} and 
 is less interesting when the objective is purely risk minimization; instead, one may focus on more sophisticated portfolios, such as those targeting a high Sharpe ratio.
In contrast,  $\one\notin \operatorname{span}(\mathbf V)$ is an analytically more interesting case. In this case, the equal-weight portfolio may have substantially larger risk than the oracle minimum risk, and  optimization is genuinely important. This case is also of empirical relevance. \cite{ding2021high}  describe the angle between $\one$ and
$\mathbf{V}^{\perp}$ as the risk diversification condition. They provide empirical evidence that in US stocks, the  decaying MVP risk is consistent with an asymptotically diversifiable risk setting---a finding also confirmed using this paper's data.    
\end{remark}

Asymptotic analysis directly on the ZVP weight $\widehat{\boldsymbol{\omega}}$ is difficult because the orthogonalization of $\mathbf S$ cannot be expressed as an easy-to-handle function of the data. 
To establish the theoretical results, it will be more convenient to work with the ridge representation in the limiting case when $\tau \to 0$ as $N,T\to \infty$.
Given the assumptions, we state the OOS variance $\mathbb V(\widehat{\boldsymbol{\omega}})$.

The expressions in Theorem \ref{t1} below
involve the \emph{Stieltjes transform}, a fundamental object in~RMT. Let $N>T$ and $\theta_1 \geq \cdots \geq  \theta_{N} > 0 $ be the eigenvalues of $\mathbf \Omega $. For $\tau \geq 0$, we define $m_{\mathbf \Omega}(-\tau)$ as the unique positive solution to 
$$
\frac{1}{m_{\mathbf \Omega}(-\tau)}=\tau+\frac{1}{T}\sum_{i=1}^N\frac{\theta_i}{1+\theta_im_{\mathbf \Omega}(-\tau)}.
$$
Based on some basic results in RMT,  we know 
$$m_{\mathbf \Omega}(-\tau)=T^{-1}{\rm tr}\{[\tau \mathbf I_T+T^{-1}\mathbf Z^\top \mathbf  \Omega \mathbf Z]^{-1}\}+O_{\rm p}(T^{-1})$$
provided that Assumptions \ref{assu} and \ref{assu:order} hold and $N/T \to \gamma \in (1, \infty)$, under which $T^{-1}\mathbf Z^\top\mathbf \Omega \mathbf Z$ is of full rank  in probability, and the eigenvalues are bounded away from zero and infinity. Then $m_{\mathbf \Omega}(0)$ is a positive bounded number. When $N/T \to \infty$, both $\|T^{-1}\mathbf Z^\top\mathbf \Omega \mathbf Z\|$ and $\|T^{-1}\mathbf Z^\top\mathbf \Omega \mathbf Z\|_{\min}$ grow at rate $N/T$ in probability, which implies $m_{\mathbf \Omega}(0) \asymp T/N$.

\begin{thm}\label{t1} 
Suppose Assumptions \ref{assu} and \ref{assu:order} hold, and $N,T\to \infty$.  
\begin{enumerate} 
\renewcommand{\labelenumi}{(\roman{enumi}).}
\item If $N/T \to \gamma \in (0,1) $, then
    \begin{equation*}\label{g54r01allmain}
    \mathbb{RV}(\widehat{\boldsymbol{\omega}})  \stackrel{{\rm p}}{\to} \frac{1}{1-\gamma }.
    \end{equation*}

\item If $N/T \to \gamma \in (1,\infty) $, then 
\begin{equation}\label{g55newr00aration1100amain}
\mathbb{V}(\widehat{\boldsymbol{\omega}})=c(0)\cdot\frac{\mathrm{tr} (\mathbf\Omega)}{N}\cdot \frac{1+O_{\rm p}\{ d(\mathbf\Omega)\}+o_{\rm p}(1)}{\mathbf 1^\top \mathbf P_{\mathbf V}^{\perp}\mathbf 1},
\end{equation}
where
$d(\mathbf\Omega)=\|[\mathbf\Omega-N^{-1}\mathrm{tr}(\mathbf\Omega)\mathbf I_N]\mathbf P_{\mathbf V}^{\perp}\|$ and
\begin{equation*}\label{tildeatau0a2main}
c(0)=\frac{1}{1-T^{-1}m_{\mathbf \Omega}^2(0)\mathrm{tr}\{[\mathbf I_N+m_{\mathbf \Omega}(0)\mathbf\Omega]^{-2}\mathbf\Omega^2\}}.
\end{equation*}
Moreover, if $\mathbf{\Omega}=\sigma^2\mathbf{I}_N$ with $\sigma^2\in (0,\infty)$, then 
\begin{equation}\label{g55newr00aration1100a1224main}
\mathbb{RV}(\widehat{\boldsymbol{\omega}})\stackrel{{\rm p}}{\to }\frac{\gamma}{\gamma-1}.
\end{equation}

\item If $N / T \to \infty$, then
    \begin{equation}
    \label{g55newr00aration11main}
    \frac{\mathbb{V}(\widehat{\boldsymbol{\omega}}) }{
    \mathbb V(\boldsymbol{\omega}_V) }\stackrel{{\rm p}}{\to } 1,
    \end{equation}
    where $\boldsymbol{\omega}_V = \mathbf P_{\mathbf V}^{\perp}\one/ ( \mathbf 1^\top\mathbf P_{\mathbf V}^{\perp}\one )$.
    Furthermore, if $\mathbf{\Omega} = \sigma^2 \mathbf I_N$, then 
    $$\mathbb{RV}(\widehat{\boldsymbol{\omega}})\stackrel{{\rm p}}{\to} 1.$$
\end{enumerate}
\end{thm}

Theorem \ref{t1}(i) states the asymptotic property of  
$\mathbb{RV}(\widehat{\boldsymbol{\omega}})$ when~\mbox{$N<T$}, where~$\mathbf S$ is of full rank.
In this regime, the relative variance $\mathbb{RV}(\widehat{\boldsymbol{\omega}})$ monotonically increases in~$\gamma$. In particular, when $\gamma$ is close to 1, the relative variance explodes. 
This instability is due to an ill-posed inverse problem near $\gamma = 1$, in the sense associated with Fredholm integral equations of the first kind; see \citet{Kress1999}.
The result also implies that $\mathbb{V}(\widehat{\boldsymbol{\omega}})$ takes a familiar U-shaped curve. 
Recall that $\mathbb{V}({\boldsymbol{\omega}}^*) =(\one^\top \boldsymbol{\Sigma}^{-1} \one)^{-1}$ decreases in $N$. When $N$ is much smaller than $T$, $\mathbb{V}(\widehat{\boldsymbol{\omega}})$ decreases as $N$ increases; this is where the first descent appears.

In the regime $N/T \to \gamma \in (1,\infty)$,  the factor $c(0) \in [1,\infty)$ in \eqref{g55newr00aration1100amain} is determined by the finite-dimensional deterministic
equivalent $m_{\bOmega}(0)$. Both quantities retain their dependence on $N$, $T$, and the
spectrum of $\bOmega$. This representation expresses the leading risk through deterministic quantities
rather than the eigenvalues of the random sample covariance matrix. A more precise asymptotic
expression is given in Theorem S.1 of the supplementary material.
(\ref{g55newr00aration1100amain}) implies that the difference between $\mathbf\Omega$ and $\mathbf I_N$ plays an important role. 
To highlight the rapid second descent after surpassing the interpolation threshold $\gamma = 1$, we present a simple expression  (\ref{g55newr00aration1100a1224main}) under the special case $\mathbf \Omega = \sigma^2 \mathbf I_N.$ 

Result (iii) is the limiting case when $N \gg T$, under which the OOS variance~$\mathbb{V}(\widehat{\boldsymbol{\omega}})$ converges to $\mathbb V(\boldsymbol{\omega}_V)$. Here 
$\boldsymbol{\omega}_V$ can be viewed as a \emph{factor-eliminating} weight, because {$\mathbf V^\top \boldsymbol{\omega}_V =\mathbf 0$} and thus 
$$
\mathbb V(\boldsymbol{\omega}_V)
= \boldsymbol{\omega}_V^\top \boldsymbol{\Sigma} \boldsymbol{\omega}_V
=\boldsymbol{\omega}_V^\top \boldsymbol{\Omega} \boldsymbol{\omega}_V.
$$
To see the implications, 
consider the case $\|\mathbf P_{\mathbf V}^{\perp} \mathbf 1\|/\sqrt N \geq c$ for some constant $c \in (0,\infty)$, under which~\mbox{$\mathbb V(\boldsymbol{\omega}_V) \asymp N^{-1}$}. Given the asymptotic equivalence,  $\mathbb{V}(\widehat{\boldsymbol{\omega}})$  leverages the advantages brought by high dimensionality when $N$ is large. 
Since 
$$\mathbb V(\boldsymbol{\omega}^*)
= (\one^\top \boldsymbol{\Sigma}^{-1} \one)^{-1} \geq (\one^\top \boldsymbol{\Omega}^{-1} \one)^{-1} \asymp N^{-1},
$$
the optimal rate is achieved by 
$\mathbb{V}(\widehat{\boldsymbol{\omega}})$.

\begin{remark}
Under the factor model, this order of $\mathbb{V}(\widehat{\boldsymbol{\omega}})$ is in general much better than that of the equal-weight (EW) portfolio. 
In fact, if $\|\mathbf P_{\mathbf V}^{\perp} \mathbf 1\| / \sqrt{N} = c N^{-\beta}$ for some $\beta\in[0,\delta/2)$ so that there is a non-trivial angle between the one-vector and the column space of $\mathbf V$,
then the OOS variance under the~EW scheme is
{
$$
\mathbb{V}( \boldsymbol{\omega}_\text{EW})
=\one^\top \mathbf \Sigma \one / N^2  \asymp N^{\delta-1}(1-c^2N^{-2\beta})+N^{-1}
$$
and
$
 N^{2\beta-1}\asymp \mathbb{V}  (\boldsymbol{\omega}_V)\asymp \mathbb{V}  (\boldsymbol{\omega}^*).
$
When $\beta>0$ or $0<c<1$, it follows
$
\mathbb{V}( \boldsymbol{\omega}_\text{EW}) \asymp  N^{\delta-1}\gg N^{2\beta-1}\asymp \mathbb{V}  (\boldsymbol{\omega}_V)\asymp \mathbb{V}  (\boldsymbol{\omega}^*).
$
This inferior OOS variance from the EW scheme is due to its disregard for the presence of the factors. 
When $c=1$ and $\beta=0$, we have $\|\mathbf P_{\mathbf V}^{\perp} \mathbf 1\| =\sqrt N$, $\|\mathbf V^\top \mathbf 1\|=0$ and
$
\mathbb{V}( \boldsymbol{\omega}_\text{EW}) \asymp  N^{-1}\asymp \mathbb{V}  (\boldsymbol{\omega}_V)\asymp \mathbb{V}  (\boldsymbol{\omega}^*),
$ 
which is a degenerate case.}
\end{remark}

Proposition \ref{prop:ridgelet} above  has shown that ZVP is the limit of ridge weight by setting the ridge penalty $\tau\to 0$. A natural question is how the performance of  ZVP compares with that of a ridge portfolio. In the case when $N\gg T$, we have the following result.
\begin{prop}\label{tauoptimal}
Suppose Assumptions \ref{assu} and \ref{assu:order} hold. If $N,T\to \infty$ and $N / T \to \infty$, then 
\begin{eqnarray*}
\frac{\inf_{\tau >0}\mathbb{V}(\widehat{\boldsymbol{\omega}}_{\tau})}{\mathbb{V}(\widehat{\boldsymbol{\omega}})}=1+o_p(1).
\end{eqnarray*} 
\end{prop}

Proposition \ref{tauoptimal} clarifies that under the specified conditions, no matter how one chooses the ridge tuning parameter, uniformly over $\tau>0$ 
the best result is asymptotically equivalent to ZVP; when $N\gg T$ choosing $\tau$ for ridge is unnecessary.
Figure \ref{opt} below provides a simulation illustration of this phenomenon.

\medskip

Notice that in general $\mathbf \Omega \neq \sigma^2 \mathbf I_N$,
and there is a gap between $\mathbb{V}(\boldsymbol{\omega}_V)$ and $\mathbb{V}({\boldsymbol{\omega}}^*)$, 
because $\boldsymbol{\Omega}$ shows up in $\boldsymbol{\omega}^*$ but not in $\boldsymbol{\omega}_V$, and 
(\ref{g55newr00aration11main}) implies that the gap persists between~$\mathbb{V}(\widehat{\boldsymbol{\omega}})$ and~$\mathbb{V}({\boldsymbol{\omega}}^*)$. 
In other words, $\mathbb{V}(\widehat{\boldsymbol{\omega}})$ does not achieve the oracle lower bound.
This is not surprising because ZVP, with a limiting ridge penalty $\tau \mathbf I_N$ as $\tau \to 0$, does not align with a general~$\boldsymbol{\Omega}$. In the special case~$\mathbf{\Omega} = \sigma^2 \mathbf I_N$, the oracle performance is attained by $\mathbb{V}(\widehat{\boldsymbol{\omega}})$, as stated in the second half of Theorem~\ref{t1}(iii).
This observation hints that, to reach the oracle OOS variance in the general case, we must utilize the information in $\boldsymbol{\Omega}$.
This is what we will do in the next section---upgrade ZVP1 to ZVP2.

\subsection{ZVP2}\label{sec:ridgelet2}

To overcome the drawback that ZVP1 ignores the covariance of the idiosyncratic returns, let us first consider an infeasible scenario for $N>T$ when $\boldsymbol{\Omega}$ is known, so we can work with a modified version of \eqref{eq:P0}:
\begin{equation*}\label{eq:P-refine}
\min_{\bw\in {\Delta}_N} \, \bw^\top \bOmega \bw  \qquad\text{subject to }
\bomega^\top \mathbf S \bomega = 0.
\end{equation*}
Again, this programming problem yields an in-sample ZVP, with the only difference being $\bOmega$ in the quadratic objective.

The corresponding ridge-penalized version replaces $\mathbf{S}_\tau = \mathbf S + \tau \mathbf I_N$ with $\mathbf{S}_{\tau,\mathbf\Omega}= \mathbf S + \tau \mathbf\Omega$, thereby producing
$$
\widetilde{\boldsymbol{\omega}}^{\mathrm{ifs}}_{\tau} = 
\frac{\mathbf{\mathbf{S}_{\tau,\mathbf \Omega}^{-1}\mathbf{1}}}
{\one^{\top}\mathbf{S}_{\tau,\mathbf \Omega}^{-1}\mathbf{1}} \rightarrow 
\frac{  \bOmega^{-1/2}\mathbf P^\perp_{\bOmega^{-1/2}\mathbf R} \bOmega^{-1/2}\one}
{\one^\top \bOmega^{-1/2}\mathbf P^\perp_{\bOmega^{-1/2}\mathbf R} \bOmega^{-1/2} \one}=\widetilde{\boldsymbol{\omega}}^{\mathrm{ifs}} \quad \mbox{ as } \tau \to 0,
$$
where the superscript ``ifs'' signifies its \emph{infeasibility}, and 
\begin{equation*}
\mathbf P^\perp_{\bOmega^{-1/2}\mathbf R}
= \mathbf{I}_N - \bOmega^{-1/2} \mathbf R(\mathbf R^\top\bOmega^{-1}\mathbf R)^{-1}\mathbf R^\top\bOmega^{-1/2}.
\end{equation*}
In practice, we will plug in an estimator $\widehat \bOmega$ for the high-dimensional $N\times N$ matrix $\boldsymbol{\Omega}$,  for example, with additional data independent of $\mathbf S$. Given $\|\bOmega\| \asymp \|\bOmega\|_{\min} \asymp 1$ in Assumption 1(i),
if an appropriate estimator satisfies $\| \widehat \bOmega\| \asymp \| \widehat \bOmega\|_{\min}  \asymp 1$ in probability, then we consider the following feasible estimator as ZVP2:
\begin{equation}\label{eq:ridgelet2_w}
\widetilde{\boldsymbol{\omega}} =
\frac{ \widehat \bOmega^{-1/2} \mathbf P^\perp_{\widehat \bOmega^{-1/2}\mathbf R} \widehat \bOmega^{-1/2}\one}
{\one^\top \widehat \bOmega^{-1/2}\mathbf P^\perp_{\widehat \bOmega^{-1/2}\mathbf R} \widehat \bOmega^{-1/2}\one},
\end{equation}
where $\bP^\perp_{\widehat \bOmega^{-1/2}\mathbf R}$ is a plug-in version of $\bP^\perp_{\bOmega^{-1/2} \mathbf R}$, replacing $\bOmega$ with  $\widehat \bOmega$. 
Obviously, the performance of ZVP2 depends on the choice of $\widehat \bOmega$. The following theorem shows that if $\widehat \bOmega$ is a high-quality estimator, then ZVP2 attains optimality.

\begin{thm}\label{t1s}

Suppose the assumptions in Theorem \ref{t1} hold, and $\widehat \bOmega$ is independent of $\bS$. As $N,T\to \infty$ and  $N / T \to \infty$:
\begin{enumerate}
\renewcommand{\labelenumi}{(\roman{enumi}).}
\item The relative variance of the infeasible estimator is 
$\mathbb{RV}(\widetilde{\boldsymbol{\omega}}^{\mathrm{ifs}}) \stackrel{{\rm p}}{\to }  1.
$ 
\item The relative variance of the feasible estimator is 
\begin{equation*}\label{g55newr00aration11newestmain}
\mathbb{RV}(\widetilde{\boldsymbol{\omega}}) =1+O_{\rm p}(\|\mathbf{ \Omega}-b\widehat {\mathbf{\Omega}}\|)+o_{\rm p}(1),
\end{equation*}
where $b=N^{-1}\mathrm{tr}(\widehat{ \mathbf{\Omega}}^{-1}\mathbf{ \Omega})$.
\item In addition, if $\| \widehat{\boldsymbol{\Omega}} - \boldsymbol{\Omega} \| = o_{\rm p}(1) $, then $\mathbb{RV}(\widetilde{\boldsymbol{\omega}})  \stackrel{{\rm p}}{\to } 1.    
$ 
\end{enumerate}
\end{thm}

Theorem \ref{t1s}(i) shows that the infeasible ZVP2, $\widetilde{\boldsymbol{\omega}}^{\mathrm{ifs}}$, achieves the oracle OOS variance and improves upon ZVP1.
Result (ii) implies that whether the feasible $\widetilde{\boldsymbol{\omega}}$ reaches the oracle OOS variance hinges on the estimation quality of $\widehat{\boldsymbol{\Omega}}$, measured by $\|\mathbf{ \Omega}-N^{-1}\mathrm{tr}(\widehat{ \mathbf{\Omega}}^{-1}\mathbf{ \Omega})\widehat {\mathbf{\Omega}}\|$. The exact representation  is  given in Equation (S.5.18)  of the supplementary material. In short, the convergence rate depends on $T/N$ and the  error in estimating $\boldsymbol{\Omega}$ in the spectral norm. For the purpose of notational ease, we omit the exact expressions without obscuring the main idea of the theorem.  

Since~$\boldsymbol{\Omega}$ is a high-dimensional covariance matrix, to consistently estimate it with limited sample size~$T$, some structural conditions must be imposed. A reasonable approach is to assume that, after removing the common factor component, the dependence among the remaining idiosyncratic components is weak. Sparsity in the idiosyncratic covariance matrix is a widely adopted condition in practice (e.g., \citealt{ding2021high,du2023high}).   
Theorem \ref{t1s}(iii) provides a verifiable condition about the estimation quality of $\widehat{\boldsymbol{\Omega}}$ for the optimality of the feasible estimator. Specifically, if $\widehat{\mathbf \Omega}$ is a consistent estimator of $\mathbf \Omega$ under the spectral norm, then $\widetilde{\boldsymbol{\omega}}$ attains the oracle OOS variance. To ensure independence, one may implement a data-splitting procedure and use a separate subsample for estimating~$\boldsymbol{\Omega}$. Accordingly, a hold-out sample is used for constructing $\mathbf{R}$. In the empirical analysis, however, we can use the full sample throughout to maximize the use of available information and mitigate the finite-sample loss induced by data splitting, particularly when $T$ is small. In the statistical literature, 
one widely used estimator that leverages idiosyncratic sparsity in high-dimensional covariance matrices is \cite{Fan_Liao_Mincheva2013}'s principal orthogonal complement thresholding~(POET). It applies adaptive thresholding on a principal component analysis (PCA)-based residual covariance matrix estimator. To make this paper self-contained, we briefly describe the~POET procedure below. 

\begin{algorithm}
\begin{tabular}{p{0.001\textwidth}p{0.1\textwidth}p{0.7\textwidth}}
\hline
\multicolumn{2}{l}{{\bf Algorithm: POET}}& 
$\quad$ {\bf Input:} Data matrix $\mathbf{R}$.
$\quad$ {\bf Output:} $\widehat{\boldsymbol{\Omega}}$
\\\hline
& Step 1.& Compute $\mathbf S=T^{-1}{\mathbf{R}}{\mathbf{R}}^{\top}$ and perform PCA on $\mathbf S$ to decompose it into  $\mathbf S=\sum_{i=1}^N\lambda_{i}\xi_i\xi_i^{\top}$, where $\lambda_{1}\geq \cdots\geq \lambda_N$ are the eigenvalues and $\{\xi_i\}_{i=1}^{N}$ are the corresponding eigenvectors. \\
&Step 2.& Estimate $r$ via the eigenvalue-ratio-based method \citep{ahn2013eigenvalue}. Set $\widehat{\mathbf{B}}=(\sqrt{\lambda_{1}}\xi_1,\ldots,\sqrt{\lambda_{r}}\xi_r)$ and $\mathbf{S}_u=(\hat{\sigma}_{ij})=\sum_{i=r+1}^N\lambda_{i}\xi_i\xi_i^{\top}$. 
\\
&Step 3.& Threshold on $\mathbf{S}_u$ to get 
$\widehat{\boldsymbol{\Omega}}=(\hat{\sigma}_{ij}^{\mathcal{T}})$ with $\hat{\sigma}_{ii}^{\mathcal{T}}=\hat{\sigma}_{ii}$ and $\hat{\sigma}_{ij}^{\mathcal{T}}=\hat{\sigma}_{ij}\cdot \mathbb{I}(|\hat{\sigma}_{ij}|\geq \mathcal{T}_{ij})$ for $i\neq j$.
Here
$\mathcal{T}_{ij}=C_1 \hat{\theta}_{ij}^{1/2}\eta_T$ with $\hat{\theta}_{ij}=T^{-1}\sum_{t=1}^T(\hat{u}_{it}\hat{u}_{jt}-\hat{\sigma}_{ij})^2$, $\eta_{T}=(r\sqrt{\log N}+r^2)/\sqrt{T}+r^3/\sqrt{N}+\sqrt{(\log N)/T}$,  $\hat{u}_{it}=((\mathbf{I}_N-\sum_{k=1}^r\xi_k\xi_k^{\top}){\mathbf{R}})_{it}$, and  $C_1>0$ is  a tuning parameter determined by 5-fold cross-validation to minimize the OOS variance for ZVP2,~$\widetilde{\boldsymbol{\omega}}$. 
\\
\hline
\end{tabular}
\end{algorithm}

For an idiosyncratic covariance matrix $\mathbf\Omega=(\sigma_{ij})$, define \mbox{$m_N=\max_{i\leq N} \sum_{j=1}^N\mathbb{I}(\sigma_{ij}\neq 0)$} as the row-wise maximum non-zero entries. \cite{Fan_Liao_Mincheva2013} show that the above idiosyncratic covariance matrix estimator satisfies  
\begin{equation}\label{1228main}
\|\widehat{\boldsymbol{\Omega}}-\boldsymbol{\Omega}\|=O_{\rm p}(m_N \eta_T).
\end{equation}
In our setting, $\eta_T =O\{\sqrt{(\log N)/T}\}$. As long as the number of row-wise non-zero entries is not too large in that $m_N \ll\sqrt{T/(\log N)} $, the POET estimator $\widehat{\mathbf \Omega}$ is consistent under the spectral norm, and  
the optimality as stated in Theorem \ref{t1s}(iii) is secured.  Since $\widehat{\Omega}$ is consistent, it is positive definite with probability tending to one; the default tuning scheme further ensures its positive definiteness for the ZVP2 weight in practice by selecting a sufficiently large tuning parameter. \cite{WangFan2017} propose
a shrinkage POET (S-POET) method which allows weak factors and corrects the estimated leading eigenvalues. The idiosyncratic covariance matrix estimated with S-POET is the same as the original POET.  \cite{WangFan2017} also show that the convergence rate of $\widehat{\boldsymbol{\Omega}}$ is unaffected by the presence of weak factors when  $\delta\in(1/2,1]$. The case $\delta\leq 1/2$ is not covered by their theoretical results.
We will employ the POET (S-POET) estimator and use the estimated idiosyncratic covariance matrix as described above in the numerical analysis.

The ZVP2 approach highlights  an
important insight that is beyond merely large covariance matrix estimation. For high-dimensional MVP optimization, it is critical to make use of the idiosyncratic covariance matrix $\boldsymbol{\Omega}$. In other words, the key to efficient MVP estimation is \emph{idiosyncratic projection}. Theorem \ref{t1s} suggests that, to attain the oracle risk, one should consistently estimate the idiosyncratic covariance matrix in {the spectral norm}. 
In comparison, shrinkage-type estimators, such as the nonlinear shrinkage estimator by \cite{ledoit2017nonlinear}, may not achieve consistency when the idiosyncratic covariance matrix is non-diagonal. Proposition S.1 in the supplementary material clarifies the out-of-sample risk of the nonlinear shrinkage based estimator under factor models with some additional conditions and $N \gg T$.

If one, on the other hand, directly constructs a plug-in weight  using a consistent large covariance matrix estimator such as the POET estimator, 
 she may not learn the MVP efficiently, due to the estimation
error from the factor component  by directly inverting the estimated full covariance matrix. For ZVP2, although  estimated factors are used to construct $\mathbf{\widehat{\Omega}}$, they serve only to remove the common component and facilitate the idiosyncratic projection. Unlike a direct plug-in portfolio based on a covariance matrix estimator, such as the POET estimator,  ZVP2 does not invert the estimated full covariance matrix, and thus avoids the direct propagation of factor estimation errors in the estimated MVP.

To be precise, when $N/T \to \gamma \in (1,\infty)$, we give an asymptotic
expression of $\mathbb{V}(\widetilde{\boldsymbol{\omega}})$ in Theorem S.5 of the supplementary material. It shows that
$$ \mathbb{V}(\widetilde{\boldsymbol{\omega}})=c_{\boldsymbol{\Omega}_2}(0)\frac{\mathbf 1^\top  \mathbf{K}_{\widehat{\boldsymbol{\Omega}}} \boldsymbol{\Omega}\mathbf{K}_{\widehat{\boldsymbol{\Omega}}}\mathbf 1}{(\mathbf 1^\top  \mathbf{K}_{\widehat{\boldsymbol{\Omega}}}\mathbf 1)^2}\left\{1+o_{\rm p}(1)\right\}$$
with
\begin{eqnarray}\label{20261013am916main}
\mathbf{K}_{\mathbf{\widehat{\Omega}}}=[\mathbf{\widehat{\Omega}}+m_1(0)\mathbf \Omega]^{-1/2}(\mathbf{I}_N-\mathbf P_{[\mathbf{\widehat{\Omega}}+m_1(0)\mathbf \Omega]^{-1/2}\mathbf V})[\mathbf{\widehat{\Omega}}+m_1(0)\mathbf \Omega]^{-1/2}.
\end{eqnarray} 
Here $c_{\boldsymbol{\Omega}_2}(0)$ is gained by replacing $\boldsymbol{\Omega}$ in $c(0)$ by $\boldsymbol{\Omega}_2=\widehat{\boldsymbol{\Omega}}^{-1/2}\boldsymbol{\Omega}\widehat{\boldsymbol{\Omega}}^{-1/2}$.
The term 
$$\mathbf 1^\top  \mathbf{K}_{\widehat{\boldsymbol{\Omega}}} \boldsymbol{\Omega}\mathbf{K}_{\widehat{\boldsymbol{\Omega}}}\mathbf 1/(\mathbf 1^\top  \mathbf{K}_{\widehat{\boldsymbol{\Omega}}}\mathbf 1)^2$$ is associated with the risk under the weight $\mathbf{K}_{\mathbf{\widehat{\Omega}}}\mathbf 1/(\mathbf 1^\top  \mathbf{K}_{\mathbf{\widehat{\Omega}}}\mathbf 1)$  due to $\mathbf{K}_{\mathbf{\widehat{\Omega}}}\mathbf V=\mathbf 0$. 
Here $m_1(0)$ is the positive solution to
\begin{eqnarray*}
\frac{1}{m_1(0)}=\frac{1}{T}\mathrm{tr}\{[\mathbf \Omega^{-1}\mathbf{\widehat{\Omega}}+m_{1}(0)\mathbf I_N]^{-1}\}.
\end{eqnarray*}
The results suggest that when $\widehat{\boldsymbol{\Omega}}$ is accurate, both ZVP2 and a direct plug-in by inverting the entire covariance matrix estimator perform well. When $\widehat{\boldsymbol{\Omega}}$ is inaccurate, however, 
$m_1(0)>0$ can be of a constant order under $N/T \to \gamma \in (1,\infty)$ and $\mathbf{K}_{\mathbf{\widehat{\Omega}}}\mathbf 1/(\mathbf 1^\top  \mathbf{K}_{\mathbf{\widehat{\Omega}}}\mathbf 1)$ retains information about the population idiosyncratic covariance $\mathbf \Omega$ through $[\mathbf{\widehat{\Omega}} +m_1(0)\mathbf \Omega]^{-1/2}$, making ZVP2 less sensitive to estimation errors in $\mathbf{\widehat{\Omega}}$ than a direct plug-in approach.
 This helps explain the robustness of ZVP2 to factor-estimation errors and achieves better MVP performance. 
In the supplementary material, we will 
demonstrate it by including the comparison in numerical examples with the POET estimator and the S-POET estimator.

\section{Simulation Studies}\label{Sec:Simu}

In this section, we conduct numerical simulations to check the quality of the asymptotic approximation. 

\subsection{Simulation Setting}\label{sec:simu-DGP}

Our return data are generated from $\mathbf{r}_{t}\sim N({\bf 0},\boldsymbol{\Sigma})$ with two configurations of $\boldsymbol{\Sigma}$:
\begin{description}
    \item[Setting 1:]   $\boldsymbol{\Sigma}=\mathbf{B}\mathbf{B}^{\top}+\sigma^2 \mathbf{I}_N$ (homoskedastic idiosyncratic returns). 
    \item[Setting 2:] $\boldsymbol{\Sigma}=\mathbf{B}\mathbf{B}^{\top}+\boldsymbol{\Omega}$, where $\boldsymbol{\Omega}$ is a sparse covariance matrix.
\end{description}
The parameters are calibrated based on empirical data of the S\&P 500 Index constituents---the last four years of test data in our empirical studies. Specifically,  we get the daily returns of the~S\&P~500~Index stocks between 2020 and 2023, which include~\mbox{$N_1=516$} stocks in total, and compute its sample covariance matrix $\tilde{\mathbf{S}}=\sum_{i=1}^{N_1}\lambda_{i}\xi_i\xi_i^{\top}$, where~\mbox{$\lambda_{1}\geq\cdots\geq\lambda_{N_1}$} and $\{\xi_i\}_{i=1}^{N_1}$ are the eigenvalues and the corresponding eigenvectors of $\tilde{\mathbf{S}}$. Based on Figure~\ref{pca_stocks}, we set~the number of factors to~$r=1$,
and thus decompose $\tilde{\mathbf{S}}=\lambda_{1}\xi_1\xi_1^{\top}+\tilde{\mathbf{S}}_u$.
We then specify the first $N_1$ rows of $\mathbf{B}$ as 
$\mathbf{B}_{1:N_1}=\sqrt{\lambda_{1}}\xi_1$. 
When $N>N_1$, e.g.,~$N=1000$, we generate the loadings for the remaining assets $\mathbf{B}_{(N_1+1): N}$, the submatrix of $\mathbf{B}$ from row $N_1+1$ to $N$, from $\text{Uniform}[b_l,b_h]$ with  $b_l$ and $b_h$ being the range of the entries of $\mathbf{B}_{1:N_1}$.

To construct the idiosyncratic covariance matrix, in Setting 1 we make  $\sigma^2=N_1^{-1}\sum_{i=1}^{N_1}(\tilde{\mathbf{S}}_u)_{ii}$.  
For Setting 2, we apply soft-thresholding to the off-diagonal elements of $\tilde{\mathbf{S}}_u$ and get a sparse idiosyncratic covariance matrix block $\boldsymbol{\Omega}_{1:N_1,1:N_1}$. 
If $N>N_1$, new diagonal entries $\diag(\boldsymbol{\Omega}_{N_1+1:N,N_1+1:N})$ are generated from $\text{Uniform}[\sigma_l,\sigma_h]$ with $\sigma_l$ and $\sigma_h$ being the range of the original $N_1$ assets. We then select  positions randomly from the remaining off-diagonal entries of~$\boldsymbol{\Omega}$ based on the original sparsity level in $\boldsymbol{\Omega}_{1:N_1,1:N_1}$ and fill them with random covariance terms. The newly assigned non-zero idiosyncratic correlations are generated from a uniform distribution with a range matching that of the first~$N_1$ assets.  Finally, we apply soft-thresholding to the off-diagonal entries of 
$\boldsymbol{\Omega}$ again to maintain the positive definiteness of the full idiosyncratic covariance matrix $\boldsymbol{\Omega}$.
{In Section S.1 of the supplementary material, we provide additional numerical analysis under other settings of covariance matrix. The results are robust. }

We simulate independent draws over $t =1,\ldots,T$ from the above models and evaluate how the portfolio performs when $N$ increases. We consider nested models with an increasing cross-sectional dimension $N=400$, $600$, $800$, and~$1000$.  
Regarding the time dimension, we focus on the performance with a small $T$ because it represents a challenging yet empirically relevant scenario. Since the environment and sentiment of the market evolve over time, it is advantageous to maintain a short investment horizon and learn investment strategies based on the most recent data, as highlighted in \cite{didisheim2023complexity}.
In both the simulation and the following empirical applications, 
we learn the portfolio weights with~$T=22$ $(44, 63)$, corresponding to one month (two months, three months) of training data with daily returns.

We simulate the data for 1000 independent replications. In each replication, we estimate the weights and evaluate the portfolio performance in the (OOS) relative risk (RR):
$$
\mathbb{RR}(\boldsymbol{\omega})=\sqrt{\mathbb{RV}(\boldsymbol{\omega})}-1.
$$
The lower bound of RR is 0. The closer this number is to 0, the better the performance.

\subsection{Illustration of Double Descent of ZVP}

In Section \ref{sec:Intro}, 
we  showcased in Figure \ref{dd_risk} the in-sample and OOS risks of ZVP1 against the number of assets~$N$  while keeping the sample size $T=22$ fixed. We used Setting 2, where the idiosyncratic covariance matrix is a sparse matrix. For brevity, we present the average results based on 10 replications, as the main pattern is already evident and remains unchanged with a larger number of replications. The double descent patterns for different settings are similar.

\begin{figure}
\begin{center}
\includegraphics[page=1,width=0.495\textwidth]
{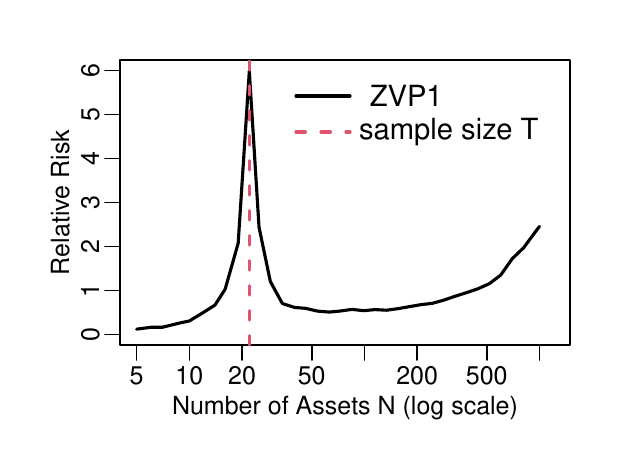}
\includegraphics[page=1,width=0.495\textwidth]
{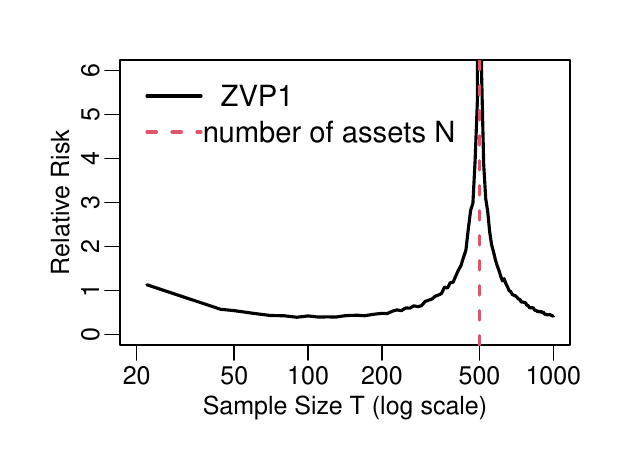}
\caption{\textbf{Relative Risk curves of ZVP1.} In the left panel, we draw the mean relative risks for various $N$s and fixed sample size $T=22$. In the right panel, we draw the mean relative risks for various $T$s and fixed dimension~$N=500$. The mean relative risk is computed as the average from 10 replications. We use  Setting 2, where the idiosyncratic covariance matrix is a sparse matrix.   }\label{risk_fixedTN50_settingIII}
\end{center}
\end{figure}

In Figure \ref{risk_fixedTN50_settingIII}, we draw the RRs of ZVP1 against two dimensions, growing $N$ while keeping the sample size fixed at $T=22$ (left panel), and growing $T$ while keeping the dimension fixed at~$N=500$ (right panel). 
In the left panel, as the number of assets $N$ increases, RR initially rises, peaking at $N=T$, and then declines. 
Towards the large cross section $N=1000$, RR goes up again. 
This does not contradict our theory which works under the asymptotics with $N,T\to \infty$; in this figure, however, $T$ is fixed while $N $ keeps increasing. 
In the right panel, we observe the curve with respect to $T$. Holding $N$ fixed, RR peaks  at~$T=N$, and climbs down in either direction as $T$ becomes smaller or greater than $N$.

Based on the messages conveyed in Figures \ref{dd_risk} and \ref{risk_fixedTN50_settingIII}, the double descent phenomenon can be explained in the following stages. 
First, in the regime~\mbox{$N<T$}, as $N$ grows from a small value, the oracle risk drops faster, dominating the growing~RR of the estimator. This leads to the first descent of the portfolio risk of ZVP. Second, when~$N$ approaches $T$, RR of the estimator increases drastically, dominating the effect of the diminishing oracle risk. This results in a peak at $N=T$. Finally, when~$N$ surpasses $T$, both RR and the oracle risk decline, leading to the second descent.

{\subsection{ZVP1 versus Optimal-Tuned Ridge}

ZVP1, as the limiting case of the ridge weight, is tuning-free. Naturally, one may ask, how  it performs compared with an optimally tuned ridge estimator. 
To illustrate, we devise an optimally tuned ridge weight estimator, which optimizes the penalization by five-fold cross-validation to minimize the out-of-sample portfolio risk. We call the resulting portfolio OptRidge. 

In Figure \ref{opt}, we draw the out-of-sample risk of the OptRidge, compared with that of the ZVP1. We use the same setting as that of Figure \ref{dd_risk}, where the number of assets $N$ grows while the sample size $T = 22$ is fixed. 
As anticipated, the optimally tuned ridge brings benefit in particular when $N$ is close to $T$ by flattening the bump of ZVP1 near the interpolation line. 

The behavior in the over-parameterized regime is interesting. The OptRidge works very similarly to ZVP1 when $N\geq 50$. Checking the chosen tuning parameter of OptRidge, we find that they are virtually zero. The results confirm that ZVP1, without tuning, behaves in the high-dimensional regime essentially the same as the optimal-tuned ridge weight.

\begin{figure}[H]
\centering{}\includegraphics[page=1,width=0.65\textwidth]{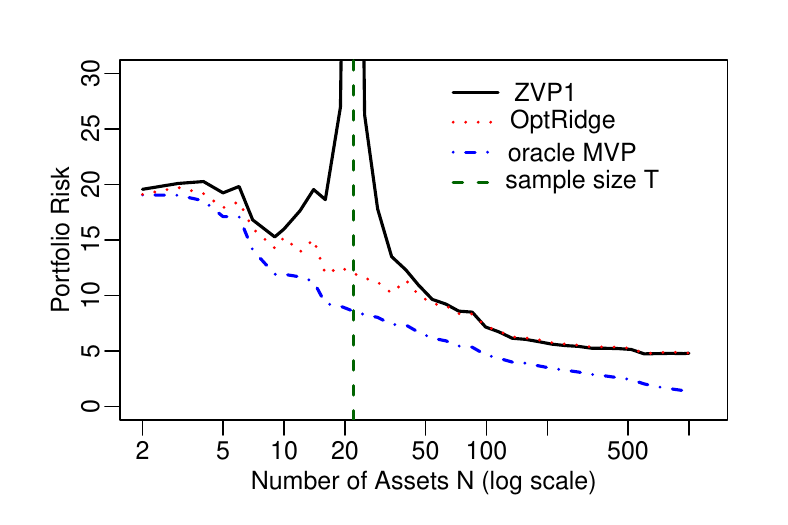}
\caption{ \textbf{Out-of-sample risk of the ZVP1 and an optimally tuned Ridge (OptRidge) portfolio.}  The optimal Ridge uses tuning parameter chosen by five-fold cross-validation with the objective of out-of-sample risk minimization.
}\label{opt}
\end{figure}

We have evaluated the risk performance of OptRidge comprehensively across all the settings we considered. We find that in terms of the risk summary statistics that we will report in subsequent sections, its performance is essentially indistinguishable from that of the ZVP1 estimator. As we focus on the challenging case when $N$ is much larger than $T$,  we omit the OptRidge result in the following tables. 
}
\subsection{Overall Performance}

We evaluate the performance of ZVP1 and ZVP2, 
in comparison with \citet{ledoit2004well}'s LS,  \citet{ledoit2017nonlinear}'s FNLS, and the EW portfolio \citep{demiguel2009optimal} as benchmarks.  
To better understand the numerical properties of ZVP2, we hold out a separate sample of size $T_0$ to estimate ${\boldsymbol{\Omega}}$. As expected, the larger  the held-out sample size, the better  the estimation quality of $\widehat{\mathbf \Omega}$. In the tables below, we will report those for $T_0=63$. 
We do not resort to the held-out sample in empirical applications in Section \ref{Sec:Emp}.

\begin{table}[tbp]

\caption{{{ \textbf{Risk performance under Setting 1.} The mean and standard deviation (in  parentheses) of RR of various portfolios from 1000 replications are reported. The smallest mean risk in each setting is shown in bold. 
}}}
\vspace{-0.6cm}
\begin{center}\small
\tabcolsep 0.1in\renewcommand{\arraystretch}{1.2} \doublerulesep
2pt
\begin{tabular}{lrccccc}
\hline\hline
&\multicolumn{6}{c}{Setting 1: $\boldsymbol{\Sigma}=\mathbf{B}\mathbf{B}^{\top}+\sigma^2\mathbf{I}_N$}   \\
  $T$&$N$&ZVP1&ZVP2&LS&FNLS&EW\\\hline
22	&400	&{0.945}	(0.396) &{\bf 0.917}	(0.369) &1.068 (0.559)	&1.002	(0.402) &\;\;7.548\\
	&600	&{1.051} (0.467)&	{\bf 1.006} (0.432)	&1.120	(0.588) &1.103 (0.473)&	11.382\\
	&800	&{1.041} (0.491)	&{\bf 0.998}	(0.451) &1.084 (0.508)&	1.086 (0.495)&	15.899\\

&	1000&	{1.099} (0.524)&	{\bf 1.040} (0.476)	&1.135	 (0.538)&1.140	 (0.527)&19.401\\\\

 44 & 400 & \;{\bf0.360} (0.131) & 0.389 (0.136) & 0.396 (0.145) & 0.381 (0.144) & \;\;7.548 \\
 & 600 & \;{\bf0.374} (0.147) & 0.398 (0.149) & 0.395 (0.154) & 0.397 (0.154) & 11.382  \\
 & 800 & \;{\bf0.354} (0.147) & 0.372 (0.148) & 0.369 (0.152) & 0.373 (0.155) & 15.899  \\
 & 1000 & \;{\bf0.366} (0.152) & 0.384 (0.161) & 0.378 (0.155) & 0.385 (0.159) & 19.401 \\\\
63 & 400 & 0.237 (0.075) & 0.275 (0.078) & 0.248 (0.083) & \;{\bf0.215} (0.087) & \;\;7.547 \\
 & 600 & 0.225 (0.085) & 0.258 (0.083) & 0.234 (0.089) & \;{\bf0.217} (0.092) & 11.382  \\
 & 800 & 0.204 (0.084) & 0.235 (0.084) & 0.210 (0.086) & \;{\bf0.201} (0.089) & 15.899  \\
 & 1000 & \;{\bf0.206} (0.088) & 0.234 (0.086) & 0.212 (0.090) & 0.207 (0.091) & 19.400  \\
\hline
\end{tabular}
\end{center}\label{simu_res_I_largeN}


\caption{{\textbf{ Risk performance of MVPs under Setting 2.} 
}}
\vspace{-0.6cm}
\begin{center}\small
\tabcolsep 0.1in\renewcommand{\arraystretch}{1.2} \doublerulesep
2pt
\begin{tabular}{lrccccc}
\hline\hline
&\multicolumn{6}{c}{Setting 2: $\boldsymbol{\Sigma}=\mathbf{B}\mathbf{B}^{\top}+\boldsymbol{\Omega}$}   \\
  $T$&$N$&ZVP1&ZVP2 & LS &FNLS &EW  \\ \hline
22 & 400  & 0.932 (0.315) & \;{\bf 0.796} (0.295) & 0.963 (0.319) & {0.803} (0.283) & \;\;6.363 \\
                    & 600  & 1.336 (0.456) & \;{\bf 1.039} (0.386) & 1.359 (0.459) & {1.085} (0.386) & \;\;9.326 \\
                    & 800  & 1.943 (0.637) & \;{\bf1.163} (0.446) & 1.968 (0.640) & 1.236 (0.459) & 12.961 \\
                    & 1000 & 2.439 (0.786) & \;{\bf1.232} (0.486) & 2.465 (0.788) & 1.327 (0.508) & 16.026 \\\\
44& 400  & 0.514 (0.106) & \;{\bf0.406} (0.114) & 0.522 (0.110) & 0.415 (0.100) & \;\;6.363 \\
                    & 600  & 0.684 (0.147) & \;{\bf0.499} (0.147) & 0.693 (0.150) & 0.536 (0.132) & \;\;9.326 \\
                    & 800  & 0.951 (0.216) & \;{\bf0.527} (0.161) & 0.962 (0.218) & 0.582 (0.151) & 12.961 \\
                    & 1000 & 1.172 (0.266) & \;{\bf0.545} (0.174) & 1.184 (0.268) & 0.600 (0.161) & 16.026 \\\\
63 & 400  & 0.431 (0.069) & 0.323 (0.066) & 0.428 (0.071) & \;{\bf 0.317} (0.065) & \;\;6.363 \\
                    & 600  & 0.539 (0.091) & \;{\bf0.362} (0.083) & 0.542 (0.093) & 0.397 (0.084) & \;\;9.326 \\
                    & 800  & 0.707 (0.128) & \;{\bf0.369} (0.093) & 0.712 (0.129) & 0.421 (0.095) & 12.961 \\
                    & 1000 & 0.847 (0.159) & \;{\bf0.371} (0.096) & 0.854 (0.161) & 0.428 (0.099) & 16.026 \\
        \hline
\end{tabular}
\end{center}\label{simu_res_III_largeN}
\end{table}

We display the RRs in Tables \ref{simu_res_I_largeN} and \ref{simu_res_III_largeN}, based on 1000 replications. 
Given the presence of  factors, in all settings EW does not work well, as it fails to diversify the risk stemming from the factors. For each fixed $N$, increasing $T$ substantially reduces the RRs of all methods except EW. For each fixed $T$, however, increasing $N$ does not necessarily reduce RR,  particularly under Setting 2. ZVP1 generally outperforms LS, while ZVP2 achieves substantially lower RRs under Setting 2.

In Setting 1, the idiosyncratic covariance matrix is exactly a multiple of the identity matrix. 
The overall best performer is ZVP2 when $T=22$ and ZVP1 when $T=44$.  When $T=22$,  the asymptotic equivalence between ZVP1 and ZVP2 may not yet be materialized in finite samples, and the PCA-based estimator of the idiosyncratic covariance matrix appears to exploit the factor-eliminating space more explicitly, benefiting ZVP2. As $T$ increases to $44$, the asymptotic regime becomes more relevant. ZVP1 converges to the factor eliminating weight that attains the oracle risk, while ZVP2 contains  additional estimation error. As a result, ZVP1 performs better.
It is also not surprising to see that, with a data-driven tuning parameter, LS cannot outperform ZVP1, a result we have theoretically shown in Proposition \ref{tauoptimal}. 
When~$T=63$,  ZVP1 maintains the edge over ZVP2, 
although the performance of our tuning-free method declines relative to FNLS since $N/T$ gets smaller.

In Table \ref{simu_res_III_largeN}, Setting 2 has a nonzero sparse idiosyncratic covariance matrix.
Here ZVP1 still generally outperforms LS.
The overall best performer is ZVP2, which takes $\mathbf \Omega$ into consideration. Its advantage is particularly salient under a large $N$, e.g., $N\geq 600$, surpassing FNLS. 
These observations corroborate our theoretical results on the asymptotic optimality of the ZVP2 estimator.

\subsection{Robustness to Heavy-tailed Distribution}

\begin{table}[!htbp]
\caption{{{ \textbf{Risk performance of MVPs under Setting 1 and heavy-tailed distribution.} Data are generated from $t_5$. The mean and standard deviation (in parentheses) of RR of various portfolios from 1000 replications are reported. The smallest mean risk in each setting is shown in bold.
}}}
\vspace{-0.6cm}
\begin{center}\small
\tabcolsep 0.1in\renewcommand{\arraystretch}{1.18} \doublerulesep
2pt
\begin{tabular}{lrccccc}
\hline\hline
&\multicolumn{6}{c}{Setting 1: $\boldsymbol{\Sigma}=\mathbf{B}\mathbf{B}^{\top}+\sigma^2\mathbf{I}_N$}   \\
$T$&$N$&ZVP1&ZVP2 & LS&FNLS&EW\\\hline

22 & 400  & 0.965 (0.410) & \;{\bf 0.909} (0.348) & 1.117 (0.489) & {0.932} (0.381) & \;\;7.548 \\
                    & 600  & 1.072 (0.479) & \;{\bf 0.995} (0.399) & 1.151 (0.516) & {1.016} (0.442) & 11.382 \\
                    & 800  & 1.069 (0.506) & \;{\bf0.981} (0.424) & 1.121 (0.527) & {1.002} (0.462) & 15.899 \\
                    & 1000 & 1.124 (0.535) & \;{\bf1.028} (0.447) & 1.167 (0.551) & {1.044} (0.485) & 19.401 \\\\
44 & 400  & \;{\bf0.362} (0.137) & 0.397 (0.132) & 0.415 (0.163) & 0.366 (0.138) & \;\;7.548 \\
                    & 600  & \;{\bf0.375} (0.155) & 0.408 (0.147) & 0.403 (0.166) & 0.379 (0.149) & 11.382 \\
                    & 800  & \;{\bf0.356} (0.159) & 0.382 (0.170) & 0.374 (0.165) & 0.360 (0.148) & 15.899 \\
                    & 1000 & \;{\bf0.368} (0.164) & 0.391 (0.151) & 0.383 (0.169) & 0.371 (0.152) & 19.401 \\\\
63 & 400  & 0.240 (0.076) & 0.284 (0.075) & 0.257 (0.090) & \;{\bf0.219} (0.083) & \;\;7.548 \\
                    & 600  & 0.226 (0.086) & 0.270 (0.083) & 0.238 (0.092) & \;{\bf0.221} (0.087) & 11.382 \\
                    & 800  & \;{\bf0.206} (0.087) & 0.249 (0.083) & 0.214 (0.090) & 0.207 (0.085) & 15.899 \\
                    & 1000 & \;{\bf 0.207} (0.090) & 0.245 (0.086) & 0.214 (0.093) & 0.212 (0.087) & 19.401 \\        \hline
\end{tabular}
\end{center}\label{simu_res_I_largeN_heavytailed}

\bigskip

\caption{{{ \textbf{Risk performance of MVPs under Setting 2 and heavy-tailed distribution.}
}}}
\vspace{-0.6cm}
\begin{center}\small
\tabcolsep 0.1in\renewcommand{\arraystretch}{1.18} \doublerulesep
2pt
\begin{tabular}{lrccccc}
\hline\hline
&\multicolumn{6}{c}{Setting 2: $\boldsymbol{\Sigma}=\mathbf{B}\mathbf{B}^{\top}+\boldsymbol{\Omega}$}   \\
$T$&$N$&ZVP1&ZVP2&LS&FNLS&EW\\\hline
22 & 400  & 0.951 (0.324) & 0.785 (0.277) & 0.997 (0.333) & \;{\bf 0.769} (0.272) & \;\;6.363 \\
                    & 600  & 1.344 (0.463) & 1.028 (0.355) & 1.382 (0.472) & \;{\bf 1.026} (0.370) & \;\;9.326 \\
                    & 800  & 1.961 (0.652) & \;{\bf 1.147} (0.417) & 2.002 (0.659) & {1.163} (0.440) & 12.961 \\
                    & 1000 & 2.462 (0.803) & \;{\bf 1.218} (0.452) & 2.507 (0.811) & {1.241} (0.484) & 16.026 \\\\
44 & 400  & 0.516 (0.110) & 0.408 (0.107) & 0.529 (0.117) & \;{\bf 0.405} (0.098) & \;\;6.363 \\
                    & 600  & 0.682 (0.157) & \;{\bf 0.501} (0.140) & 0.697 (0.162) & 0.518 (0.131) & \;\;9.326 \\
                    & 800  & 0.952 (0.227) & \;{\bf 0.530} (0.159) & 0.971 (0.232) & 0.560 (0.151) & 12.961 \\
                    & 1000 & 1.177 (0.284) & \;{\bf 0.545} (0.164) & 1.199 (0.289) & 0.577 (0.161) & 16.026 \\\\
63 & 400  & 0.429 (0.066) & 0.327 (0.064) & 0.428 (0.079) & \;{\bf 0.315} (0.058) & \;\;6.363 \\
                    & 600  & 0.540 (0.090) & \;{\bf 0.366} (0.081) & 0.544 (0.095) & 0.394 (0.078) & \;\;9.326 \\
                    & 800  & 0.711 (0.130) & \;{\bf 0.374} (0.091) & 0.720 (0.134) & 0.416 (0.089) & 12.961 \\
                    & 1000 & 0.850 (0.161) & \;{\bf 0.376} (0.095) & 0.863 (0.165) & 0.423 (0.095) & 16.026 \\        \hline
\end{tabular}
\end{center}\label{simu_res_III_largeN_heavytailed}
\end{table}

The simulation results in Tables \ref{simu_res_I_largeN} and \ref{simu_res_III_largeN} are based on data that follow the normal distribution. Financial data, however, often exhibit heavy-tailedness. 
Though heavy-tailed distributions are not covered by our Theorems \ref{t1} and \ref{t1s} for the convenience of the theoretical development, we use simulations here to check how the performance is affected by heavy-tailed distributions.

To investigate this,  we generate $\mathbf{R} =  \boldsymbol{\Sigma}^{1/2}\mathbf{W}$,  where each entry of $\mathbf{W}$ is i.i.d. standardized~$t$-distributed with the degree-of-freedom being $5$, denoted by~$t_5$. We adopt the same covariance matrix settings as before and
report the RRs in Tables \ref{simu_res_I_largeN_heavytailed} and \ref{simu_res_III_largeN_heavytailed}.
ZVP1 and ZVP2 still perform well and generate lower risks than the benchmarks in various scenarios. The results demonstrate the robustness of our proposal to heavy-tailed distributions. 

We observe that when $T=63$ (the maximum $T$ in the simulations), ZVP2
exhibits an opposite trend in Setting 1 (Tables \ref{simu_res_I_largeN} and \ref{simu_res_I_largeN_heavytailed}) and Setting 2 (Tables \ref{simu_res_III_largeN} and \ref{simu_res_III_largeN_heavytailed}). Specifically, its performance generally improves as $N$  increases under Setting 1, while it deteriorates as 
$N$  increases under Setting 2.
A key reason stems from (\ref{1228main}): under Setting 1, $m_N$ is always equal to 1, whereas under Setting 2, $m_N$ increases with $N$, thereby leading to an inflated error.

\section{Empirical Applications}\label{Sec:Emp}

According to the population formulation, diversification is a ``free lunch'' that reduces the risk in the financial market. 
Given a large number of investable assets, we must devise a scheme to enjoy the free lunch. 
In our real data applications, we first apply ZVP to the S\&P 500 Index constituents. 
Next, we extend the scope to multinational investment opportunities by incorporating the Japanese Nikkei 225 data. 

\subsection{S\&P 500 Index Stocks}

\begin{figure}
\begin{center}
\includegraphics[page=1,width=0.49\textwidth]{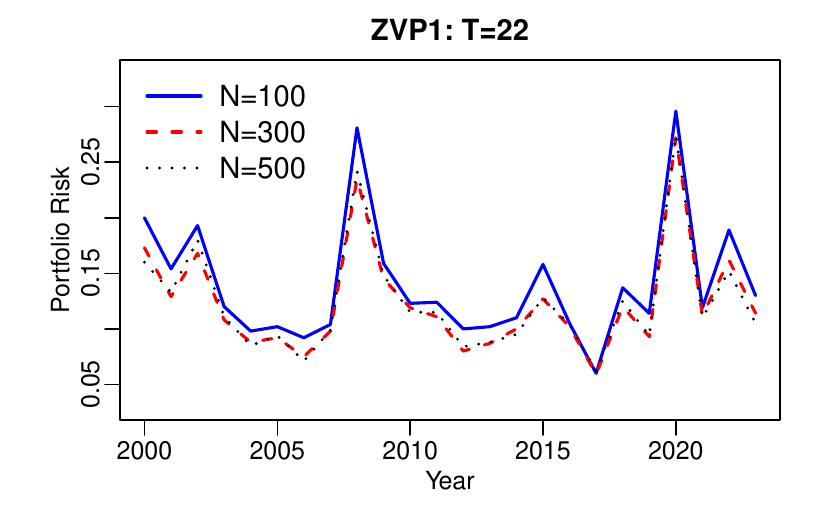}
\includegraphics[page=3,width=0.49\textwidth]{Figs/riskby_year_dimension_ZVP}
\includegraphics[page=4,width=0.49\textwidth]{Figs/riskby_year_dimension_ZVP}
\includegraphics[page=6,width=0.49\textwidth]{Figs/riskby_year_dimension_ZVP}
\caption{\textbf{Time-series of risks of MVPs between 2000 and 2023,  learned with different numbers of stocks.} We draw the annualized risks of MVP for $N=100$, $300$ and $500$ using the largest stocks by their capitalization. We include the  ZVP1 (top) and ZVP2 (bottom) portfolios for training window $T=22$ and $T=63$.  }\label{yearlyrisk_dimension}
\end{center}
\end{figure}

We collect the daily returns of the S\&P 500 stocks between January 1999 and December~2023. At the beginning of each month, we use the
historical returns of the constituent stocks for the past~$T$ days to estimate the MVP weights. We use the stocks that possess complete observations in the training window and construct portfolios with sample sizes~$T\in \{22, 44, 63\}$ days, as in the simulations.  We track the OOS daily returns of the portfolio for one month. We re-estimate the weights every month. 

We first fix a training window $T$ and examine how the risk by ZVP1 or ZVP2 varies with the number of stocks $N$. To do so,  we select the largest $N=100$, $300$, and $500$ stocks in the~\mbox{S\&P~500 Index}, sorted by their capitalization, and construct MVPs with these nested subsets of stocks.  
{
Due to the small $T$ in our empirical applications, we use the same training window to estimate $\widehat \bOmega$ and $\mathbf S$ in ZVP2, and the results are competitive.}
Figure~\ref{yearlyrisk_dimension} presents the annualized risk profiles of the ZVP1 and ZVP2 portfolios. We observe a notable reduction in risks of both ZVP1 and ZVP2 when $N$ grows from $100$ to~$300$.
When $N$ increases from $300$ to $500$, the improvement is small for ZVP1, while that for ZVP2 is still visible, in particular following the two peaks in the aftermath of the 2008 financial crisis and~2020 Covid depression.

The overall risk is reported in Panel A of Table \ref{emp}. 
The pseudoinverse estimator does not work well and even underperforms EW with small $T$. 
All other data-driven methods---ZVP1, ZVP2, LS and FNLS---clearly beat EW. It implies that when we are able to track the constituent stocks, EW is not a competitive alternative. 
While ZVP1 and LS explore the factor structure, ZVP2 and FNLS further utilize the information in the idiosyncratic component to attain risk reduction. 
ZVP2 performs better than FNLS by a small margin. 

It appears that ZVP1 aligns with LS. In the real data, the \emph{bona fide} estimator of~LS produces a tuning parameter $\tau$ so small that it approaches the limiting ridge weight. 
On the other hand, ZVP2 performs similarly to FNLS.
Theoretically optimal among the \emph{rotation-equivariant}-type estimators, FNLS is known to be one of the strongest estimators for the S\&P 500 (\citealt{ledoit2017nonlinear,ding2021high,jiang2024dynamic}). 
The equally strong performance of ZVP2 reveals the gain from utilizing the sparsity of the covariance of the idiosyncratic component. 
Our ZVP2 is a flexible framework that allows users to plug in their preferred consistent estimator to explore the structure of the problems under investigation.

\begin{table}[h]
\caption{{{ \textbf{Risk performance of MVPs over January 2000 to December 2023.} The reported quantities are the annualized risk. In Panel A, we construct the MVP portfolio using daily returns of S\&P 500 Index constituent stocks.  In Panel B, we construct multinational portfolios using both S\&P 500 Index stocks and Nikkei 225 Index stocks. The lowest risk in each setting is highlighted in bold. 
}}}
\begin{center}
\tabcolsep 0.14in\renewcommand{\arraystretch}{1.45} \doublerulesep
2pt
\begin{tabular}{lcccccc}
\hline\hline
 Training window   & ZVP1 &ZVP2 & LS  &FNLS &EW&{Pseudoinverse}\\\hline
& \multicolumn{6}{l}{Panel A: S\&P 500 Stocks}\\
$T=22$  &0.133	&{\bf 0.097}	&0.135	&0.099 &0.213 &0.226	\\
 $T=44$  &0.125	&{\bf 0.094} &	0.126	&0.096 &	0.213 &0.274\\
   $T=63$  &0.122  &{\bf 0.098}& 0.122  &0.099 &0.213&0.155\\
        \hline
          &\multicolumn{5}{l}{Panel B: S\&P 500  plus Nikkei 225 Stocks}\\
         $T=22$  &0.115	&{\bf 0.084}&	0.114&0.087 &	0.176&0.318	\\
  $T=44$ &0.108	&{\bf 0.081}&	0.108	&	0.083&0.174&0.179 \\
  $T=63$ &	0.103&{\bf 0.081}&	0.104	&	0.083&0.175 &0.251\\\hline
\end{tabular}
\end{center}\label{emp}
\end{table}

\subsection{Multinational Portfolio: S\&P 500 plus Nikkei 225}

To harvest the blessing of dimensionality, we augment the U.S.~data with Japanese data. Specifically, we build multinational portfolios by enlarging the pool of the S\&P 500 Index stocks with the Nikkei~225 Index constituent stocks. 

The number of stocks is hence brought up to about $N=725$. 
In this exercise, we follow the same estimation time framework as that used for the~S\&P 500 data. At each estimation point, we include the same S\&P 500 stocks as  in Panel A of Table \ref{emp}; in addition, we incorporate the Nikkei 225 Index stocks that have at least 75\% complete observations within the training window. As a result, the total number of involved stocks varies slightly over different sample sizes~$T$, 
thereby resulting in the small variation in EW.

We summarize the total risks of the multinational investment in Panel B of Table~\ref{emp}. 
Except for the pseudoinverse estimator, the most salient observation is that, in every case with an enlarged pool of assets, the corresponding risks of all portfolios are reduced. 
Compared with the case of investing solely in the S\&P 500 stocks, the risk of ZVP2 for the multinational pool drops from 0.097 to 0.084 with a training window $T=22$. Moreover,  
ZVP2 consistently outperforms both ZVP1 and LS, and it maintains the small edge over FNLS. It exhibits robustness across varying sample sizes and dimensionalities, making it a strong candidate for high-dimensional portfolio optimization with limited data.

The analysis above focuses on the portfolio's risk, the main target of the paper. 
In Section S.2 of the supplementary material, we also report the portfolio's performance in the mean return, Sharpe ratio and  turnover. The proposed methods do not incur a high turnover. Moreover, ZVP2 achieves a reasonable return performance in terms of its Sharpe ratio, in particular with a large $N$.

\section{Conclusion}\label{Sec:Conc}

This paper studies the generalizability of the ZVP on test data. We clarify that when~$N>T$, the pseudoinverse estimator with $\mathbf S^+$ is not a proper estimator for the MVP problem. 
We derive the exact solution of the minimum $\ell_2$-norm ZVP.
To understand the asymptotic behavior of ZVP, we use RMT to analyze the limiting behavior of ridge in a high-dimensional regime where the number of stocks $N$ can be much larger than the size of the training sample~$T$.
In an approximate factor model, the theoretical analysis motivates a refinement by incorporating the estimator of the idiosyncratic covariance matrix  $\boldsymbol{\Omega}$. We show that when $\boldsymbol{\Omega}$ can be estimated consistently, the refined ZVP2 attains the oracle OOS variance. 
Extensive simulations and empirical applications to the S\&P 500 Index constituents and the Nikkei 225 stocks demonstrate the competitive performance of the ZVP portfolios.

\bigskip
\bigskip

\appendixtitleon
\appendixtitletocon

\begin{appendices}

\section{}\label{Proofs}
\subsection{Proof of Lemma \ref{lem:existence-of-zero-var}}

Because $\mathbf S=T^{-1}\mathbf R\mathbf R^\top$, we have
$
\boldsymbol{\omega}^\top\mathbf S\boldsymbol{\omega}
=T^{-1}\|\mathbf R^\top\boldsymbol{\omega}\|^2.
$
Thus, zero in-sample variance is equivalent to
$\boldsymbol{\omega}\in\ker(\mathbf R^\top)=\ker(\mathbf S)$.
Let $\mathbf v=\mathbf P_S^\perp\one$. Since
$\ker(\mathbf R^\top)^\perp=\mathrm{span}(\mathbf R)$, the assumption
$\one\notin\mathrm{span}(\mathbf R)$ implies $\mathbf v\neq\mathbf0$.
Moreover, for every $\boldsymbol{\omega}\in\ker(\mathbf R^\top)$,
orthogonality gives
$
\one^\top\boldsymbol{\omega}
=(\mathbf P_S^\perp\one)^\top\boldsymbol{\omega}
=\mathbf v^\top\boldsymbol{\omega}.
$
Consequently, the feasible set in \eqref{eq:P0} is precisely
$\{\boldsymbol{\omega}\in\ker(\mathbf R^\top):
\mathbf v^\top\boldsymbol{\omega}=1\}$, which is a non-empty set because $\mathbf v/\|\mathbf v\|^2$ is in it.

For any feasible $\boldsymbol{\omega}$, the Cauchy-Schwarz inequality yields
$
1=\mathbf v^\top\boldsymbol{\omega}
\leq \|\mathbf v\|\|\boldsymbol{\omega}\|.
$
Equality holds if and only if $\boldsymbol{\omega}$ is proportional to
$\mathbf v$; the constraint then determines the proportionality
constant uniquely. Hence the unique minimum-norm feasible vector is given by 
$
{\mathbf v}/\|\mathbf v\|^2
=\mathbf P_S^\perp\one/
(\one^\top\mathbf P_S^\perp\one)
=\widehat{\boldsymbol{\omega}}.
$

\subsection{Proof of Proposition \ref{prop:ridgelet}}

Since $\mathbf P_S^\perp=\mathbf I_N-\mathbf U_T\mathbf U_T^\top$, we have 
$
\bS_\tau
=\mathbf S+\tau\mathbf I_N 
=\mathbf U_T\mathbf\Lambda_T\mathbf U_T^\top
+\tau(\mathbf U_T\mathbf U_T^\top+\mathbf P_S^\perp) 
=\mathbf U_T(\mathbf\Lambda_T+\tau\mathbf I_T)\mathbf U_T^\top
+\tau\mathbf P_S^\perp$. 
The two summands act on orthogonal subspaces because
$\mathbf U_T^\top\mathbf P_S^\perp=\mathbf0$ and
$\mathbf P_S^\perp\mathbf U_T=\mathbf0$. Therefore, using
$\mathbf U_T^\top\mathbf U_T=\mathbf I_T$ and
$(\mathbf P_S^\perp)^2=\mathbf P_S^\perp$, direct multiplication verifies
$
[\mathbf U_T(\mathbf\Lambda_T+\tau\mathbf I_T)\mathbf U_T^\top
+\tau\mathbf P_S^\perp]
[\mathbf U_T(\mathbf\Lambda_T+\tau\mathbf I_T)^{-1}\mathbf U_T^\top
+\tau^{-1}\mathbf P_S^\perp]  =\mathbf U_T\mathbf U_T^\top+\mathbf P_S^\perp
=\mathbf I_N$. 
This proves \eqref{eq:smw-ridge}. Substituting that identity into
\eqref{eq:ridgelet_w} and multiplying the numerator and denominator by
$\tau$ gives \eqref{Ridgelet_repres}.

It remains to take the limit. Since
$\mathbf P_S^\perp\one\neq\mathbf0$ and 
$
\one^\top\mathbf P_S^\perp\one>0,$ the limit representation follows as $\tau \to 0$ eliminates the other terms in \eqref{Ridgelet_repres}, which gives 
$
\widehat{\bomega}_\tau \to 
(\mathbf P_S^\perp\one)/(\one^\top\mathbf P_S^\perp\one)
=\widehat{\boldsymbol{\omega}}
$
as claimed.

\end{appendices}

\bigskip
\bigskip

\bibliographystyle{apalike} 
\bibliography{RMT_factor}

\includepdf[pages=-]{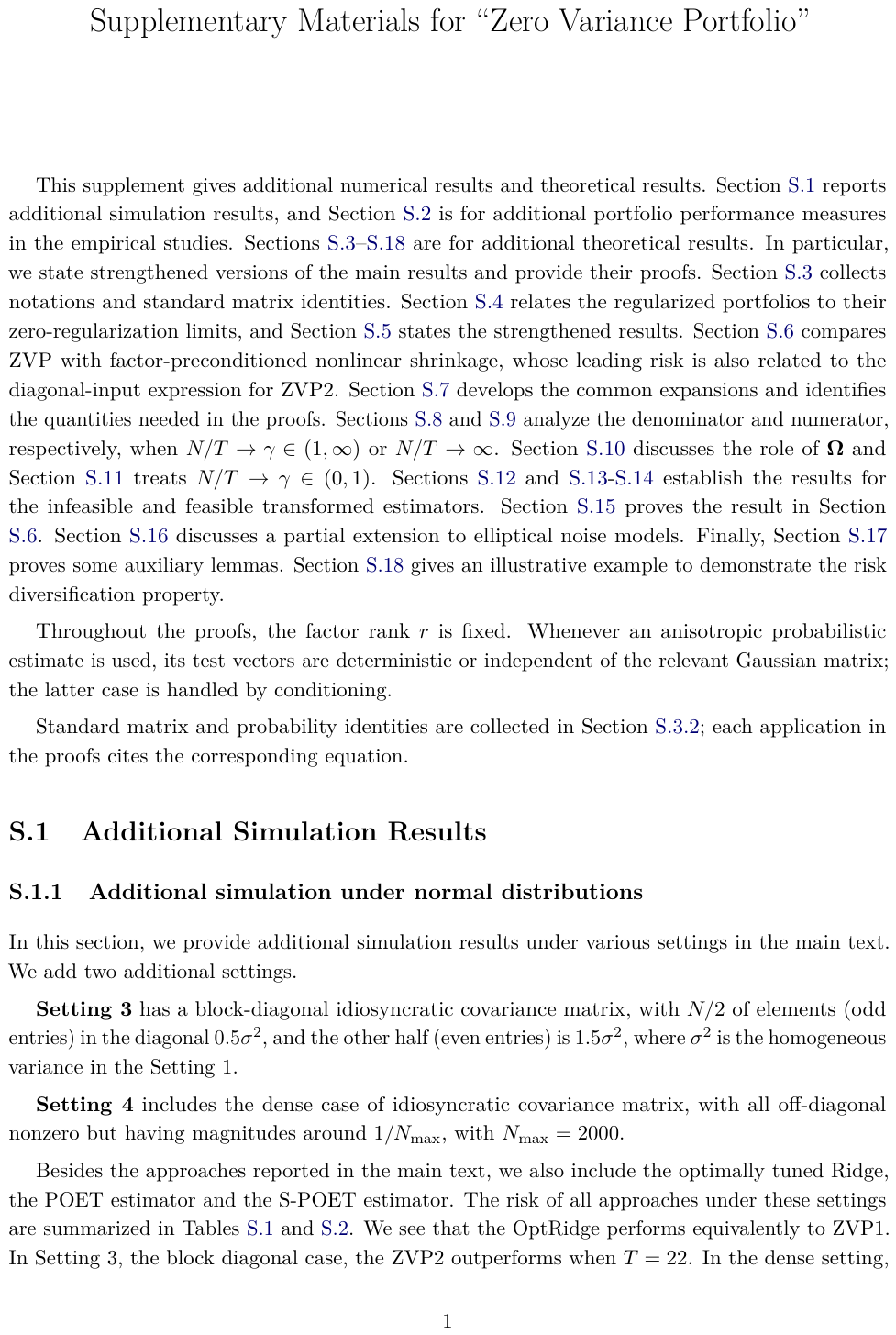}
\end{document}